\documentclass[12pt]{article}
\usepackage{amsmath,amssymb,bm,epsfig,afterpage}
\usepackage{cite}

\renewcommand{\thefootnote}{\fnsymbol{footnote}}

\begin{document}

\title{
\begin{flushright}
\begin{minipage}{0.2\linewidth}
\normalsize
WU-HEP-14-03 \\*[50pt]
\end{minipage}
\end{flushright}
{\Large \bf 
The 126 GeV Higgs boson mass and naturalness in (deflected) mirage mediation
\\*[20pt]}}

\author{Hiroyuki~Abe\footnote{
E-mail address: abe@waseda.jp} \ and \ 
Junichiro~Kawamura\footnote{
E-mail address: junichiro-k@ruri.waseda.jp}\\*[20pt]
{\it \normalsize 
Department of Physics, Waseda University, 
Tokyo 169-8555, Japan} \\*[50pt]}

\date{
\centerline{\small \bf Abstract}
\begin{minipage}{0.9\linewidth}
\medskip 
\medskip 
\small
We study the mass of the lightest CP-even Higgs boson
in the deflected mirage mediation that is a quite general framework of the mediation of supersymmetry breaking, incorporating the case where all of the modulus-, the anomaly- and the gauge-mediated 
contributions to the soft supersymmetry breaking parameters become sizable.
We evaluate the degree of tuning the so-called $\mu$ parameter required for realizing 
a correct electroweak symmetry breaking 
and study how to accomplish both the observed Higgs boson mass and the relaxed fine-tuning.
We identify the parameter space favored from such a perspective 
and show the superparticle mass spectrum with some input parameters inside the indicated region.
The results here would be useful when we aim to prove the communication 
between the visible and the hidden sectors in supergravity and superstring models based on the recent observations.
\end{minipage}
}

\begin{titlepage}
\maketitle
\thispagestyle{empty}
\clearpage
\tableofcontents
\thispagestyle{empty}
\end{titlepage}

\renewcommand{\thefootnote}{\arabic{footnote}}
\setcounter{footnote}{0}

\section{Introduction}
\label{intro}
The low-energy supersymmetry is one of the well-motivated and still promising candidate for 
the physics beyond the standard model (SM), 
which is now being tested by many experiments including the Large Hadron Collider (LHC) 
and the cosmological observations.
In particular, the minimal extension of the SM, so-called minimal supersymmetric standard model (MSSM), 
is very attractive due to the absence of quadratic divergence, 
existence of dark matter candidate, gauge coupling unification 
and radiative electroweak (EW) symmetry breaking (For a review, see e.g. Ref.~\cite{Martin:1997ns}).
The MSSM is also motivated by some ultraviolet completions of the SM such as supergravity and superstring 
models of elementary particles, 
where the most free parameters in the MSSM would be determined by the structure of the vacuum.

The recent experimental results at the LHC exclude light supersymmetric partners of the SM particles 
(called superparticles or sparticles) 
and the allowed region of the mass of the lightest CP-even Higgs boson is 
in between 124.4 and 126.8 GeV~\cite{Aad:2012tfa,Chatrchyan:2012ufa}.
The latter result implies that top squark mass (a supersymmetric partner of the top quark) 
is larger than 10 TeV or the left-right mixing of the top squarks is sizable.
However heavy top squarks bring the fine-tuning problem to the MSSM.
The mass of $Z$ boson is related to the peculiar combinations of MSSM parameters 
in the condition for triggering a successful EW symmetry breaking, such that
\begin{align}
m_Z^2 & \simeq -2 \left| \mu(M_{\rm EW}) \right|^2 -2 m_{H_u}^2(M_{\rm EW}), 
\label{EWSB}
\end{align}      
where $\mu(M_{\rm EW})$ and $m_{H_u}(M_{\rm EW})$ 
are a supersymmetric higgsino mass parameter 
and a soft supersymmetry breaking mass for the up-type Higgs boson at the EW scale, respectively.
Since $m_Z \sim 91.2\ \rm GeV$ is observed, 
the fine-tuning is required between $\mu$ and $m_{H_u}$ if these values are significantly larger than $m_Z$.

Heavy top squarks generally induce a large $|m_{H_u}|$ through the renormalization group (RG) running 
due to a large top Yukawa coupling.
Therefore, the mass of top squark should be small in order to avoid the fine-tuning problem mentioned above,
and then a sizable top squark mixing is necessary to realize the allowed Higgs boson mass without the fine-tuning.
As pointed out in Refs.~\cite{Abe:2007kf,Abe:2012xm}, 
such a desired situation can appear with the certain gaugino mass ratio 
at the so-called Grand Unified Theory (GUT) scale $M_{\rm GUT} \sim 2 \times 10^{16} \rm{GeV}$ 
where the values of the three SM gauge coupling constants unify.
Particularly, a moderately large ratio between the wino and the gluino mass 
helps both to realize the observed large Higgs boson mass 
and to relax the degree of tuning the $\mu$ parameter in order to satisfy the condition~(\ref{EWSB}).

The supersymmetry breaking is one of the most important phenomenological ingredient 
for supersymmetric models.
The supertrace theorem tells  
that we can obtain the supersymmetry breaking parameters consistent with current experimental results 
if and only if the supersymmetry breaking takes place in a sequestered sector, the so-called hidden sector.
Then, the values of soft supersymmetry breaking parameters absolutely depend on how to mediate
the supersymmetry breaking from the hidden sector to the MSSM (visible) sector. 
There are basically three kinds of mediation mechanisms those are known as 
the gravity mediation~\cite{Hall:1983iz, Soni:1983rm, Brignole:1997dp}, 
the gauge mediation~\cite{Giudice:1998bp} 
and the anomaly mediation~\cite{Randall:1998uk, Giudice:1998xp, Bagger:1999rd}.
In these cases, supersymmetry breaking is mediated at the tree-level 
by gravitational (nonrenormalizable) interactions, 
at the loop-level by gauge interactions and by a super-Weyl anomaly, respectively. 
In the bottom-up approach beyond the SM, 
phenomenologies brought by these mediation mechanisms have been analyzed in detail 
with the assumption that the contribution from one of them 
dominates those from the other mediation mechanisms.

In the top-down approach to the physics beyond the SM, 
depending on the situation, 
it sill sometimes occurs 
that more than one type of the above three mediation mechanisms give sizable contributions 
simultaneously to the soft supersymmetry breaking parameters.
For instance, the modulus (gravity) mediation and the anomaly mediation contribute comparably 
to the soft parameters in the so-called KKLT-type moduli stabilization mechanism~\cite{Kachru:2003aw} 
and this mixed mediation is recognized 
as the mirage mediation~\cite{Choi:2004sx, Choi:2005ge, Endo:2005uy,Choi:2005uz,Choi:2006xb}.
It is remarkable that the RG running of the soft parameters is compensated 
by the effects of anomaly mediation if certain presumable conditions are satisfied 
within the framework of mirage mediation.
As a result, the sparticle masses tend to unify at some lower scale 
below the GUT scale in a typical case of mirage mediation,
and from a phenomenological perspective, 
this behavior makes it easier to control the values of soft parameters 
near the EW scale~\cite{Choi:2006xb, Kitano:2005wc}.

In the mirage mediation, the ratio between the wino to the gluino mass parameter at the GUT scale 
is moderately large 
if the anomaly-mediated contribution is twice larger than the one from the (tree-level) modulus mediation 
with a suitable normalization.
A lot of phenomenological analyses have been done for this attractive case, 
that is, the so-called TeV scale mirage mediation~\cite{
Choi:2006im,Cho:2007fg,Nagai:2007ud,Nakamura:2008ey} 
where the gluino and the wino masses are unified at the TeV scale. 
It is shown in Refs.~\cite{Asano:2012sv,Kobayashi:2012ee} 
that the Higgs boson mass can reach the observed value, 
while the collect EW symmetry breaking can occur without the fine-tuning 
in the next-to-minimal supersymmetric standard model, 
the so-called NMSSM\footnote{See, e.g., Ref.~[26] for a review of the NMSSM.},  
with the TeV scale mirage mediation. 
Therefore the mirage mediation framework could be preferred from both the observed large Higgs boson mass 
and the naturalness argument, 
since a relatively large wino mass helps to satisfy both of them.

The more general framework referred to as the deflected mirage mediation is studied 
in Refs.~\cite{Everett:2008ey,Everett:2008qy}
where the gauge mediation also contributes comparably to the soft parameters 
as well as those from the other two mediation mechanisms.
Such a generalized mediation mechanism can be constructed 
by adding the mediator field $X$ and the messenger fields $\Psi ,\overline{\Psi}$ 
as an usual gauge mediation scenario.
Furthermore, it is suggested that a certain stabilization mechanism for the mediator field can explain 
how the size of gauge-mediated contribution becomes comparable with the other two mediations.
Some phenomenological aspects of the deflected mirage mediation were studied in Refs.~\cite{Choi:2009jn,Holmes:2009mx,Altunkaynak:2010xe,Altunkaynak:2010tn} .

In this paper, we aim to identify the region in the parameter space of the (deflected) mirage mediation, 
where both the experimentally allowed Higgs boson mass 
and the relaxed fine-tuning of the $\mu$ parameter are realized, 
that restricts the mediation mechanism 
and would reveal the detailed connection between the visible and the hidden sector in supersymmetric models. 

The following sections are organized as follows. 
In Section~\ref{theory}, we mention about the theoretical backgrounds 
which achieve comparable contributions from more than one mediation mechanisms.
The analytical formulae for the soft parameters adopted in the later sections are also described in this section.
In Section~\ref{pheno}, we explain the guideline for the analysis in this paper 
and the experimental bounds we take into account.
In Section~\ref{result}, we perform numerical analyses 
and identify the mediation mechanism with desired properties by specifying the model parameters.
Finally, we conclude this paper in Section~\ref{conclusion}.
In Appendix~A, we show the numerical values of Yukawa couplings adopted in the analysis for concreteness, 
though they do not play essential roles in the conclusion of this paper.

\section{Theoretical background}
\label{theory}
In this section, we review the most general framework of the mediation mechanism of supersymmetry breaking, 
that is, the deflected mirage mediation~\cite{Everett:2008ey,Everett:2008qy}, 
based on a model shown in Ref.~\cite{Everett:2008ey}.
Let us start with the four-dimensional $\mathcal{N} = 1$ effective supergravity description 
of the KKLT-type models.
When $T$, $X$ and $\Phi_i$ denote a K\"{a}hler modulus of the internal space, 
a SM gauge singlet and the MSSM matters respectively, 
the K\"{a}hler potential at the leading order has the form
\begin{align}
K=-3 \log (T+\overline{T}) + \frac{X\overline{X}}{(T+\overline{T})^{n_X}} + \frac{\Phi_i \overline{\Phi}_i}{(T+\overline{T})^{n_i}}, 
\label{kahler}
\end{align}
where $n_X$ and $n_i$ are the modular weights of $X$ and $\Phi_i$ respectively, 
those describe their profiles in the internal space.
Note that we take the unit such that the Planck mass $M_p=2.4\times10^{18}\ \rm{GeV}$ is unity.

The superpotential is assumed to be
\begin{eqnarray}
W = W_0(T) +W_1(X)+\lambda X \Psi \overline{\Psi} + W_{\rm MSSM}, 
\label{super}
\end{eqnarray}
where $W_0(T)$ and $W_1(X)$ are responsible for stabilizing $T$ and $X$, respectively.
The $N_{\rm mess}$ pairs of messenger fields $(\Psi, \bar{\Psi})$ are 
$({\bm 5}, \bm{\bar{5}})$ representations of $SU(5)$ respectively as usual gauge mediation models.
The MSSM superpotential $W_{\rm MSSM}$ contains the Yukawa interaction terms 
and the supersymmetric Higgs mass term referred to as the $\mu$ term.  

Furthermore, the (universal) profiles of the MSSM gauge fields are assumed in the internal space yielding 
gauge kinetic functions in the following form:
\begin{equation}
\label{gkf}
f_a(M_{\rm GUT})=T.
\end{equation}
Here and hereafter, $a=1,2,3$ label the gauge groups of the MSSM, $U(1)_Y,\ SU(2)_L,\ SU(3)_C$, respectively.

From the above setup, we can compute the soft parameters for the MSSM matters
contained in the soft supersymmetry breaking Lagrangian,
\begin{equation}
-\mathcal{L}_{\rm soft}=\phi^{*i}{{m^2}_i}^j\phi_j + \left[ \frac{1}{2} M_a\lambda^a \lambda^a
+ a^{ijk}\phi_i\phi_j\phi_k + \rm{h.c.} \right] ,
\end{equation}
where ${{m^2}_i}^j$, $M_a$ and $a^{ijk}$ are the scalar mass parameters, the gaugino mass parameters 
and the scalar trilinear couplings respectively.

In the deflected mirage mediation~\cite{Everett:2008ey,Everett:2008qy}, 
the messenger scale $M_{\rm mess} \equiv \lambda \langle X \rangle$
at an intermediate scale is assumed following the standard gauge mediation models. 
Thus the soft parameters at the GUT scale are identical with those of the pure mirage mediation 
and can be calculated with the method proposed in Ref.~\cite{Giudice:1997ni} as follows:
\begin{align}
M_a(M_{\rm GUT}) &= \frac{F^T}{T+\overline{T}}+ \frac{g^2_0}{16\pi^2}b'_a \frac{F^C}{C}, \\
a^{ijk}(M_{\rm GUT}) &=(3-n_i-n_j-n_k)\frac{F^T}{T+\overline{T}}
      -\frac{1}{16\pi^2}\big[y^{ljk}{\gamma_l}^i + (i \leftrightarrow j)+(i \leftrightarrow k)\big] 
            \frac{F^C}{C}, \\
{{m^2}_i}^j(M_{\rm GUT}) &= (1-n_i)\left|\frac{F^T}{T+\overline{T}}\right|^2 {\delta_i}^j
     -\frac{{{\theta}_i}^j}{16\pi^2} \left( \frac{F^T}{T+\overline{T}} \frac{F^{\overline{C}}}{\overline{C}} 
 +\rm{h.c.} \right) -\frac{{\dot{\gamma}_i}^j}{(16\pi^2)^2}\left|\frac{F^C}{C}\right|^2 ,
\label{baresoft}
\end{align} 
where the modular weights $n_i$, $n_j$, $n_k$ 
generally take different values depending on the profiles of the MSSM matter
contents labeled by $i,\ j,\ k$ and we assume the flavor-independent modular weights 
until Subsection~\ref{Fdmirage}.
In these expressions, $g_0$ is the unified gauge coupling at the GUT scale 
and $b'_a \equiv b_a + N_{\rm mess} \quad (a=1,2,3)$ represent the beta functions for the gauge couplings above the messenger threshold scale.
Thereby, $b_a$ correspond to the beta functions for the gauge couplings of the pure MSSM 
and then $(b_1,b_2,b_3)=(\frac{33}{5}, 1, -3)$ for $U(1)_Y$, $SU(2)_L$ and $SU(3)_C$, respectively.
The explicit forms of the anomalous dimensions $\gamma$ and their derivatives $\theta$, $\dot{\gamma}$ can be expressed as
\begin{align}
{{\gamma}_i}^j&=\sum_a 2c_a(\Phi_i )g_a^2{{\delta}_i}^j-\frac{1}{2}\sum_{l,m}y_{ilm}y^{jlm}, \\
{{\theta}_i}^j&=\sum_a 2c_a(\Phi_i )g_a^2{{\delta}_i}^j-\frac{1}{2}\sum_{l,m}(3-n_i-n_l-n_m)y_{ilm} y^{jlm}, \\
{{\dot{\gamma}}_i}^j&=\sum_a 2c_a(\Phi_i )b'_a g_a^4{{\delta}_i}^j
        -\frac{1}{4}\sum_{l,m} \left( {b_y}_{ilm}y^{jlm}+y_{ilm} {b_y}^{jlm} \right),
\label{anomalous} 
\end{align} 
where ${b_y}^{ijk}$ represents the beta function for the Yukawa coupling $y^{ijk}$and 
$y_{ijk}\equiv(y^{ijk})^*, {b_y}_{ijk}\equiv ({b_{y}}^{ijk})^*$.

The messengers are decoupled below the messenger scale
and then the threshold corrections to soft parameters should be taken into account at the scale.       
These contribute to the gaugino and the scalar mass parameters, while the trilinear couplings are not affected.
The messenger threshold corrections to the gaugino masses and the soft scalar mass matrices 
can be obtained by the same way as shown in Ref.~\cite{Chacko:2002et},
\begin{align}
\Delta M_a(M_{\rm mess}) &= -N_{\rm mess} \frac{g_a^2(M_{\rm mess})}{16\pi^2} 
                                    \left( \frac{F^C}{C}+\frac{F^X}{X} \right), \\
\Delta {{m^2}_i}^j(M_{\rm mess}) 
    &= \sum_a 2c_a(\Phi_i )N_{\rm mess} \frac{g_a^4(M_{\rm mess})}{(16\pi^2)^2}
     \left( \left| \frac{F^C}{C} \right|^2 + \left| \frac{F^X}{X} \right|^2 
             + \frac{F^C}{C}\frac{F^{\overline{C}}}{\overline{C}}\right){{\delta}_i}^j,
\label{barethre}
\end{align}      
where $c_a(\Phi_i)$ is the quadratic Casimir for the matter field $\Phi_i$.        
Note that all these parameters are defined in the field basis on which the kinetic terms are canonically normalized.

It is convenient to parameterize the magnitudes of the anomaly and the gauge mediated contributions
with respect to their ratio to that of modulus mediated ones. 
When the magnitude of modulus mediation is represented by $m_0\equiv F^T/(T+\overline{T})$ 
which describes the overall scale of the soft parameters, contributions 
from the other two mediations are expressed as~\cite{Everett:2008ey,Everett:2008qy}
\begin{align}
\frac{F^C}{C}&=m_0\alpha_m \ln \frac{M_p}{m_{3/2}}, \\
\frac{F^X}{X}&=\alpha_g\frac{F^C}{C}=m_0\alpha_g \alpha_m\ln \frac{M_p}{m_{3/2}},
\end{align}
where $m_{3/2}$ denotes the gravitino mass.
With this parametrization, the soft parameter formulae at the GUT scale (\ref{baresoft}) are rewritten as
\begin{align}
M_a(M_{\rm GUT}) &= m_0\left[1
          + \frac{g^2_0}{16\pi^2}b'_a \alpha_m \ln \frac{M_p}{m_{3/2}} \right] , \label{gaugino} \\
a^{ijk}(M_{\rm GUT}) &=m_0\left[ (3-n_i-n_j-n_k)
      -\frac{1}{16\pi^2}\big[y^{ljk}{\gamma_l}^i + (i \leftrightarrow j)+(i \leftrightarrow k)\big] 
            \alpha_m \ln \frac{M_p}{m_{3/2}} \right] , \label{aterm} \\ 
{{m^2}_i}^j(M_{\rm GUT}) &= {m_0}^2 \left[ (1-n_i){\delta_i}^j
     -\frac{2{{\theta}_i}^j}{16\pi^2} \alpha_m \ln \frac{M_p}{m_{3/2}}
        -\frac{{\dot{\gamma}_i}^j}{(16\pi^2)^2}\left(\alpha_m \ln \frac{M_p}{m_{3/2}}\right)^2 \right]. \label{smass}
\end{align} 
Similarly, the threshold corrections at the messenger scale become
\begin{align}
\Delta M_a(M_{\rm mess}) &= -m_0 N \frac{g_a^2(M_{\rm mess})}{16\pi^2} 
                                    \alpha_m (1+ \alpha_g) \ln \frac{M_p}{m_{3/2}}, \label{gauginoth}\\
\Delta {{m^2}_i}^j(M_{\rm mess}) 
    &= {m_0}^2\sum_a 2c_a(\Phi_i )N \frac{g_a^4(M_{\rm mess})}{(16\pi^2)^2}
     \left[ \alpha_m (1+ \alpha_g) \ln \frac{M_p}{m_{3/2}}\right]^2{{\delta}_i}^j. \label{smassth}
\end{align}  

Let us comment on the (rational) values of the parameters $\alpha_m$ and $\alpha_g$.
The former $\alpha_m$ is identical to the $\alpha$ parameter in the pure mirage mediation adopted 
in Ref.~\cite{Choi:2005uz}.
The original KKLT model~\cite{Kachru:2003aw} predicts $\alpha_m=1$, 
and many other KKLT-type models suggest $\alpha_m\sim\mathcal{O}(1)$~\cite{Abe:2005rx,Abe:2005pi}. 
In particular, models predicting $\alpha_m\simeq 2$ are fascinating from the phenomenological 
viewpoint~\cite{Kitano:2005wc,Choi:2005uz,Endo:2005uy,Choi:2006xb}, 
because the size of the Higgs soft mass parameter $m_{H_u}$ in Eq.(\ref{EWSB}) 
can become the same order of magnitude as the EW scale 
as a consequence of the RG evolution.
This ameliorates the degree of tuning the $\mu$ parameter to bring a successful EW symmetry breaking 
even if the overall scale of soft parameters are considerably larger than the EW scale.

This phenomena can be interpreted as follows.
In the mirage mediation, gaugino masses are unified at some energy scale, 
a so-called mirage scale $M_{\rm mirage}$, typically lower than the GUT scale 
and they are related by
\begin{align}
M_{\rm mirage} = M_{\rm GUT} \left( \frac{m_{3/2}}{M_{\rm P}} \right) ^{\frac{\alpha_m}{2}} .
\label{miruni}
\end{align}
This implies that $M_{\rm mirage}$ is around a TeV scale if $\alpha_m \sim 2$.
Moreover, soft masses and A-terms are also unified at $M_{\rm mirage}$ 
if Yukawa couplings are negligible in their RG evolutions 
or modular weights for the fields feeling sizable Yukawa couplings $y^{ijk}$ satisfy the condition 
\begin{equation}
\sum_{l=i,j,k} (1-n_l) = 1.
\label{unification}
\end{equation}
In this case, the values of soft parameters are highly controllable ~\cite{Kitano:2005wc,Choi:2006xb} 
including $m_{H_u}$ near the EW scale, 
that can lead $m_{H_u}\sim m_Z$ and then $\mu \sim m_Z$ through the condition (\ref{EWSB}), 
so the tuning of $\mu$ parameter is relaxed.

This attractive feature was first derived with the condition 
that the all three  types of soft parameters are unified respectively at the mirage scale~\cite{Choi:2005uz}.
However the mirage mediation has the tendency to relax the tuning of $\mu$ parameter 
even if the condition (\ref{unification}) is not satisfied any more,
which is indicated by the argument about the relation 
between the gaugino mass ratio at the GUT scale and the degree of tunning the $\mu$ parameter.
As pointed out in Refs.~\cite{Abe:2007kf,Abe:2012xm}, 
a moderately large ratio of wino to gluino mass parameters at the GUT scale is essential 
for relaxing the tuning of $\mu$ parameter through the RG running, 
and actually $\alpha_m \sim 2$ corresponds to the most desired ratio of wino to gluino mass parameter 
with the gauge kinetic function (\ref{gkf}).  
Therefore the most important ingredient to relax the fine-tuning is only the mirage unification 
of wino and gluino masses at the TeV scale,
and then we don't need to stick to the unification of soft parameters other than these two. 
For this reason, 
we take the several patterns of modular weights independently to the condition (\ref{unification}) in this paper.

It is remarkable that the mirage unification of gaugino masses do also occur 
in the deflected mirage mediation~\cite{Everett:2008qy}.
The deflected mirage scale can be written as
\begin{align}
M_{\rm mirage} = M_{\rm GUT} \left( \frac{m_{3/2}}{M_{\rm P}} \right) ^{\frac{\alpha_m \rho }{2}} ,
\end{align}
where $\rho$ is defined as
\begin{align}
\rho = \frac{1+\frac{2N_{\rm mess}  {g_0}^2}{16\pi^2} \ln \frac{M_{\rm GUT}}{M_{\rm mess}}}
                                   {1- \alpha_m \alpha_g \frac{N_{\rm mess} {g_0}^2}{16\pi^2} \ln \frac{M_{P}}{m_{3/2}}}.
\end{align}
Although it depends on several parameters, 
the gaugino masses are unified at the deflected mirage scale 
which is again lower than the GUT scale typically and it can be taken to a TeV scale.

The parameter $\alpha_g$ can also take various values of $\mathcal{O}(1)$ depending 
on the stabilization mechanism for the singlet field $X$.
As an example, the following form of the superpotential can stabilize $X$
\begin{align}
W_1(X)=\frac{X^n}{\Lambda^{n-3}},
\end{align}
with $n \ge 3$ (higher order stabilization) or $n < 0$ (nonperturbative stabilization).
In these cases, the ratio of gauge to anomaly mediation becomes 
$\alpha_g = -2 / (n-1)$~\cite{Everett:2008ey, Pomarol:1999ie}.
On the other hand, a radiative potential can stabilize $X$ and $\alpha_g=-1$ is obtained 
even when the tree-level superpotential for $X$ is absent.

From the above observations, $\alpha_m$ and $\alpha_g$ take various values of  $\mathcal{O}(1)$ 
depending on the detailed setups.
Besides, these values are determined by the moduli stabilization mechanism 
and they would depend on only the discrete parameters like, e.g.,  
the winding number of D-branes, 
the number of fluxes that generate moduli potential, 
the power of the uplifting potential 
or the above superpotential $W_1(X)$ and so on. 
Thus we treat $\alpha_m$ and $\alpha_g$ as free parameters of $\mathcal{O}(1)$ 
and they are assumed to be fixed to the values 
with an enough accuracy depending on the moduli stabilization mechanisms.
In addition to these two ratio parameters, there are free parameters in the deflected mirage mediation, 
those are overall mass scale $m_0$ of the soft parameters, 
the messenger scale $M_{\rm mess}$, 
the number of $5$ and $\bar{5}$ representation messenger pairs $N_{\rm mess}$ 
and the modular weights $n_i$, 
as well as the ratio of two Higgs vacuum expectation values (VEV) $\tan\beta$ 
and the sign of $\mu$ parameter.

\section{Phenomenological background}
\label{pheno}
In this section, we mention about a relation between the mass of the lightest CP-even Higgs boson 
and the degree of tuning the $\mu$ parameter.

\subsection{The Higgs boson mass and the tuning of $\mu$ parameter}
The degree of tuning the $\mu$ parameter is one of the most significant theoretical guideline 
to probe the physics behind the MSSM, 
because it would describe a certain naturalness of the observed EW symmetry breaking 
caused radiatively by a supersymmetry breaking effects 
whose origin is mostly independent to that of the $\mu$ term. 
There is no reason why the $\mu$ parameter has the almost equal value 
to the (particular combination of) the soft supersymmetry breaking parameters.
While we can anticipate some mechanisms determine the values of soft parameters, 
e.g. gaugino masses, and even their ratios with the required accuracy 
such as the moduli stabilization mechanisms in the corresponding string model for example.
Then some desired relations or even cancellations would be expected among the soft parameters, 
in contrast to the one between the $\mu$ parameter 
and the soft parameters which have the different origins from each other.
Therefore the tuning of $\mu$ parameter is the most serious one from this perspective. 
We define the degree of tuning the $\mu$ parameter as 
\begin{align}
\Delta_\mu=\left|\frac{\partial \ln m_Z^2}{\partial \ln {\mu_0}^2}\right|,
\end{align}
where $\mu_0$ is the initial value of the $\mu$ parameter at the GUT scale. 
We call $\Delta_\mu^{-1} \times 100 \%$ degree of tuning the $\mu$-parameter~\cite{Barbieri:1987fn}.

In this paper, we include the effect of full 1-loop MSSM RG evolution to the EW scale from the GUT scale 
with the full components of Yukawa matrices (whose numerical values are exhibited in 
Appendix~\ref{yukawa} for concreteness).
Then using the obtained values of the soft parameters near the EW scale, 
we evaluate the Higgs boson mass $m_h$ based on 
the RG improved 1-loop effective potential including the (s)top and (s)bottom contributions 
derived in Ref.~\cite{Carena:1995wu}, 
where we include the RG effects by solving the RG equations numerically 
and do not adopt the leading log approximation 
in order to assure the enough numerical precision.

\subsection{Model parameters}
For the input parameters in the MSSM, 
we take the sign of $\mu$ as positive and $\tan\beta = 15$ 
to obtain a tree-level SM-like Higgs boson mass as large as possible.
As for those parameters peculiar to the deflected mirage mediation 
we mainly study the dependences of the Higgs boson mass, the degree of tuning $\mu$ 
and the other mass spectrum on the ratio parameters $\alpha_m$ and $\alpha_g$ 
with the fixed values of modular weights ${n_i}$, the overall scale for the soft parameters $m_0$, 
the number of messenger fields $N_{\rm mess}$ and the messenger scale $M_{\rm mess}$.

First, we analyze the case with $N_{\rm mess}=0$, 
namely, pure mirage mediation with several sets of modular weights 
assigned commonly to all the quark/lepton supermultiplets 
but differently, if necessary, to the two Higgs supermultiplets respectively throughout this paper. 
Second, we treat the case of indeed deflected mirage mediation 
and examine its dependence on the property of the messenger sector.
In the deflected mirage mediation, the number of messengers $N_{\rm mess}$ are restricted 
by a condition that the Landau poles of all the gauge couplings are absent up to the GUT scale 
depending on the messenger scale $M_{\rm mess}$. 
Finally, we consider the case 
that the  supersymmetry-breaking mediations are dependent to the generation of squarks and sleptons, 
especially, its contribution to the first and the second generation is larger than that to the third generation.
In this case, several phenomenological advantages are found as we will see later, 
though certain flavor-dependent structures 
of the supersymmetry-breaking and its mediation sectors are required in the UV completion 
(which is beyond the scope of this paper) in order to obtain such a situation.

\subsection{Experimental bounds}
The recent results from the search for Higgs boson and supersymmetric particles 
at the LHC put experimental bounds on the masses of these particles.
The mass of the SM-like Higgs boson $m_h$ must reside in the range between 124.4 and 126.8 GeV, 
that requires a large radiative correction to the Higgs boson mass in the MSSM.
A lot of analyses based on simplified models have been studied and these give somewhat stringent bounds 
especially for the colored superparticles~\cite{ATLAS:007}.
The mass of the lightest top squark $m_{\tilde{t}_1}$ must be heavier than 
about 700 GeV if the neutralino LSP is lighter than 300 GeV.
For other colored sparticles, the degenerate mass of the first and the second generation squarks $m_{\tilde{q}}$ 
less than 1.8 TeV and the gluino mass $m_{\tilde{g}}$ less than 1.6 TeV are excluded when these are comparable. 
However these stringent constraints are based on such an assumption 
that the other sparticles are decoupled except the neutralino LSP 
whose mass is less than 400 GeV.
The gluino mass bound 
which is significant for the naturalness argument is relaxed as $m_{\tilde{g}} \gtrsim 1.4\ {\rm TeV}$ 
when the first and the second generation squarks are also decoupled.

\section{The Higgs boson mass, naturalness and sparticle spectra}
\label{result}
In this section, we search the parameter space of the (deflected) mirage mediation 
and identify the region allowed by the current experimental data, 
especially the observed Higgs boson mass, 
and also measure the degree of tuning the $\mu$ parameter in such a region.

\subsection{Pure mirage mediation}
First we analyze the case with $N_{\rm mess}=0$, namely, the pure mirage mediation. 
We examine the low-energy mass spectra with several patterns of modular weights, 
specifically their influence on the Higgs boson mass.
We assume modular weights have universal values 
for all the matter fields except Higgs fields among each generation,
those are denoted for the matter and Higgs fields by $n_Q$ and $n_H$ respectively.

Naively speaking, 
the magnitude of top squark $A$-term $A_t$ becomes relatively large compared with those of top squark masses 
due to such the form of soft terms coming from modulus mediation as shown in Eqs.~(\ref{aterm}) and (\ref{smass}) 
if the modular weights of the scalar quarks $n_q$ and $n_u$ are small. 
This would cause a larger Higgs boson mass than usual due to the large left-right mixing of top squarks.
The values of modular weights for the Higgs multiplets affect not only the size of $A_t$ 
but also the magnitudes of the soft Higgs mass parameters $m_{H_u}$ and $m_{H_d}$.
As a result, these influence the Higgs potential at the low-energy 
or equivalently the property of EW symmetry breaking.

\begin{figure}
\centering
\hfill
\includegraphics[width=0.45\linewidth]{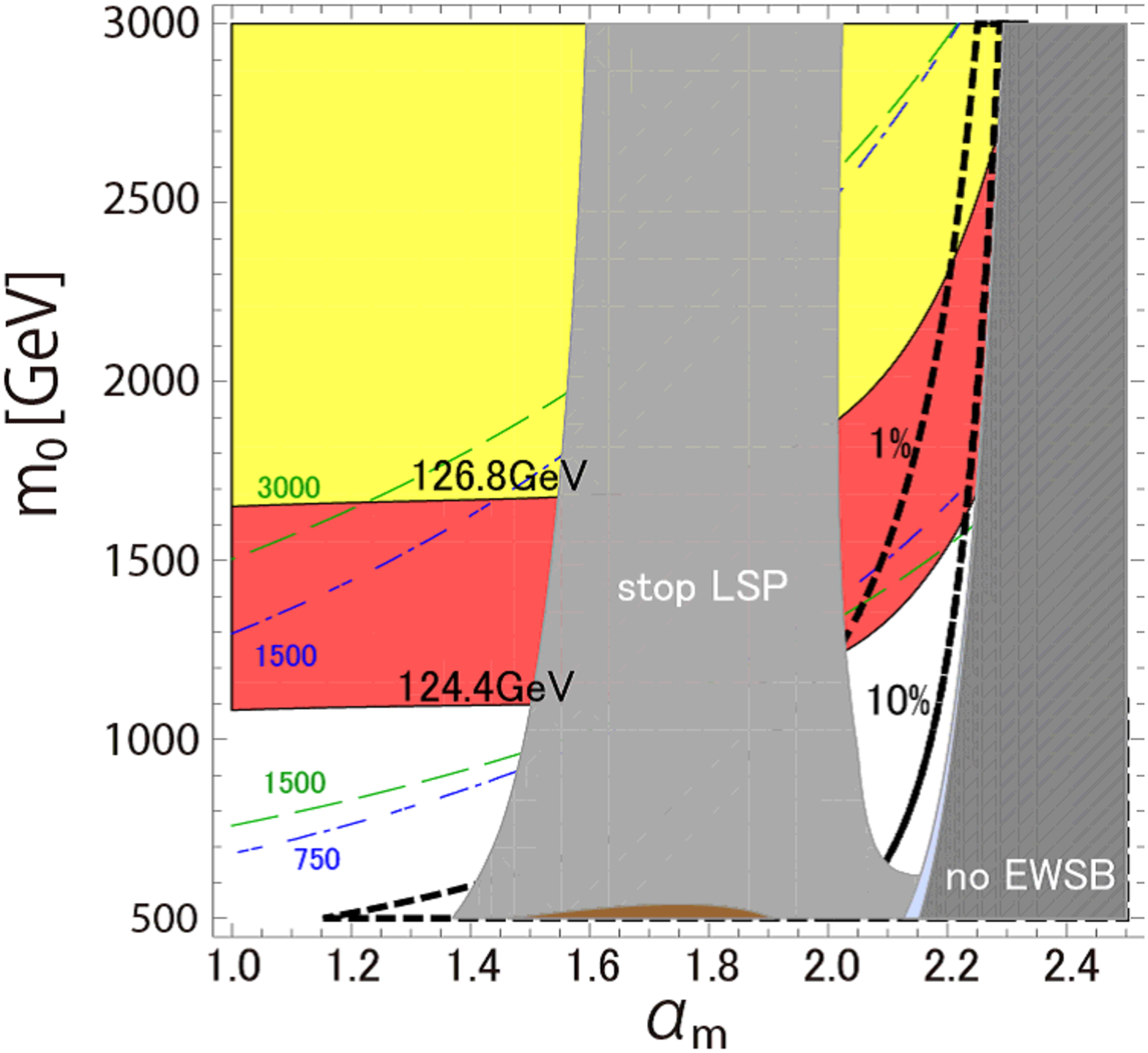} 
\hfill 
\includegraphics[width=0.45\linewidth]{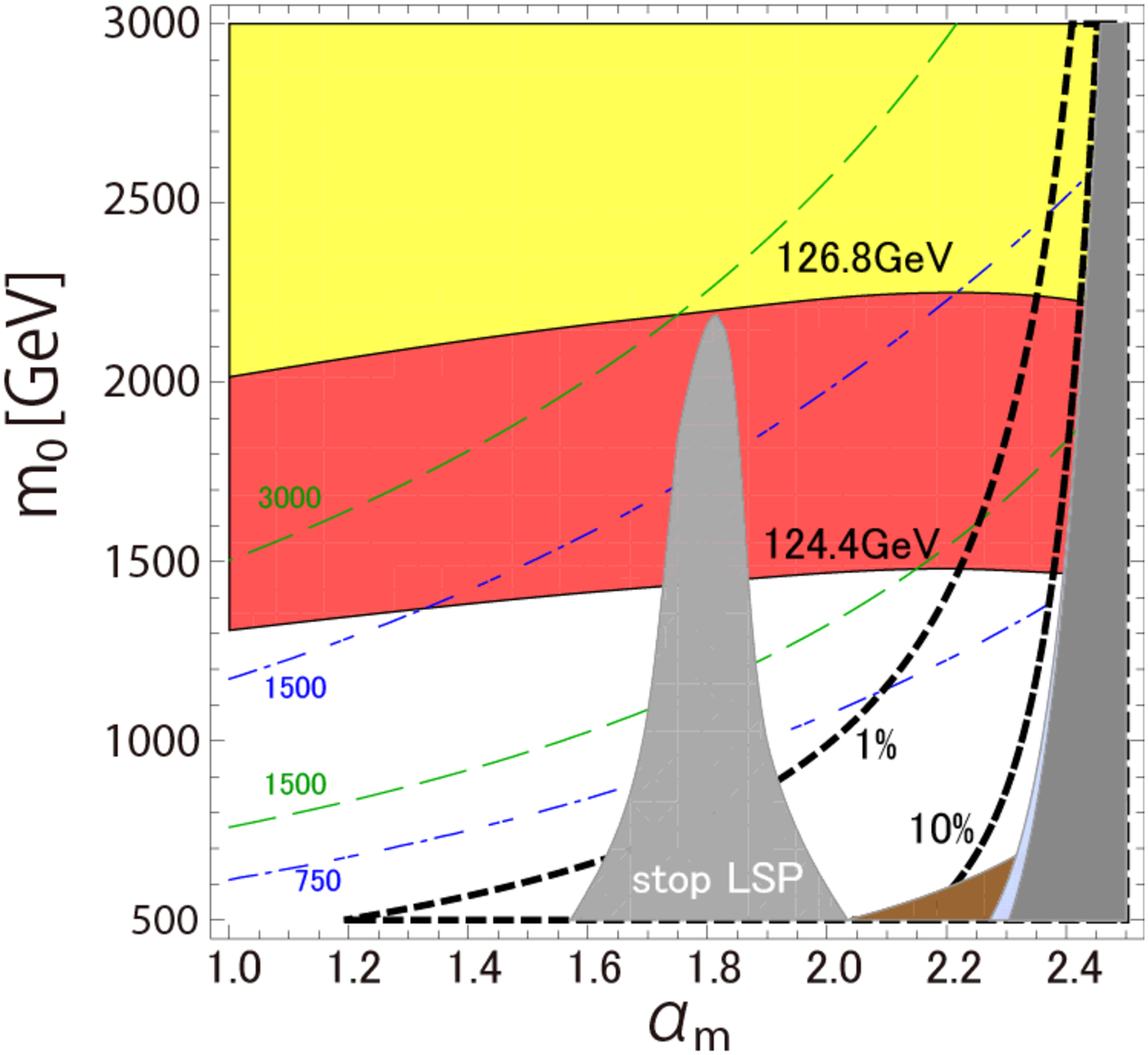}
\hfill 
\caption{Contours of the fixed Higgs boson mass $m_h$ in the unit of GeV 
and the degree of tuning $\mu$,  $|\Delta_\mu|^{-1} \times 100 \%$,  
in the pure mirage mediation on $\alpha_m$-$m_0$ plane 
with the modular weights $(n_Q, n_H) = (0, 0)$ (left panel) and $(0, 0.5)$ (right panel).
The meanings of each lines and colored regions are explained 
in the corresponding paragraphs referring to this figure. }
\label{higgsM12}
\end{figure}

\begin{figure}
\centering
\hfill
\includegraphics[width=0.45\linewidth]{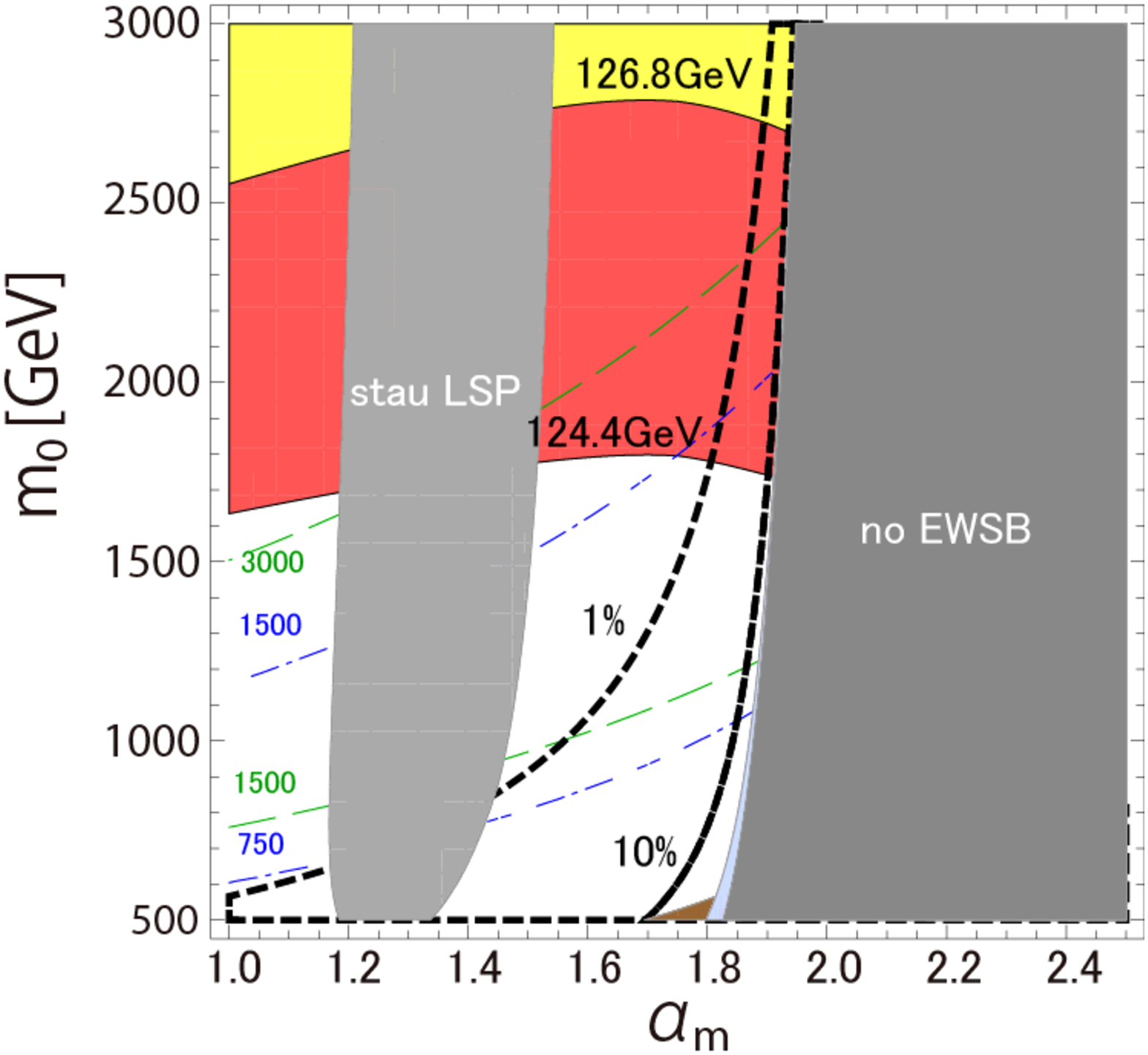}
\hfill 
\includegraphics[width=0.45\linewidth]{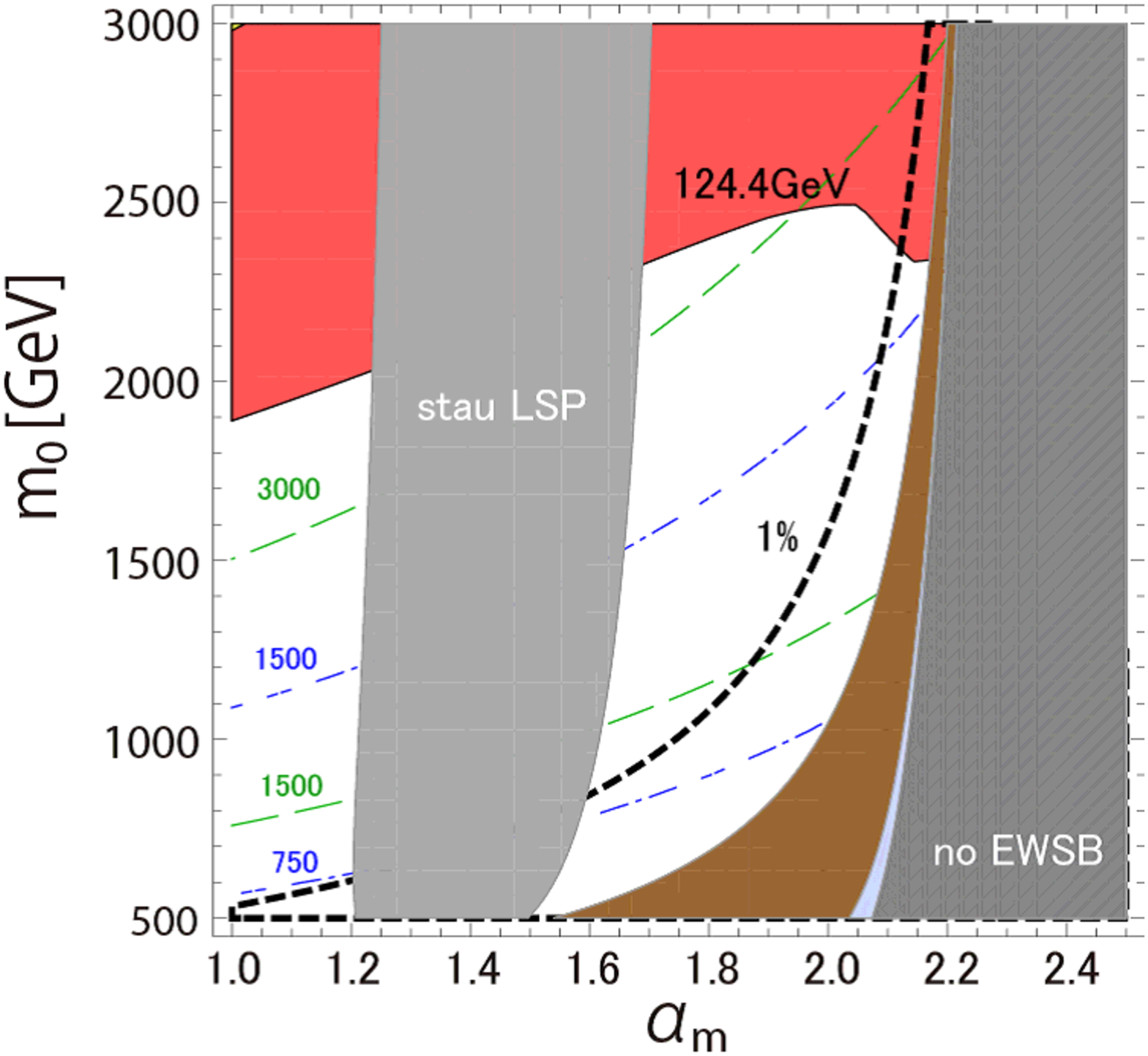}
\hfill
\caption{Contours of the Higgs boson mass and the degree of tuning in the pure mirage mediation on $\alpha_m$-$m_0$ plane with the modular weights $(n_Q, n_H) = (0.5, 0.5)$ (left panel) and $(0.5, 1)$ (right panel). 
The lines and colored regions are drawn in the same way as those in Fig.~1.}
\label{higgsM34}
\end{figure}

Figs.~\ref{higgsM12} and \ref{higgsM34} depict the Higgs boson mass $m_h$ 
and degree of tuning $\mu$ parameter $|\Delta_\mu| \times 100 (\%)$ on the $\alpha_m$-$m_0$ plane 
in the case of pure mirage mediation, 
where $\alpha_m$ is the ratio between anomaly and modulus mediated contributions to soft parameters 
and $m_0$ is the size of modulus mediation.
Figs.~\ref{higgsM12} and \ref{higgsM34} are drawn 
with the different pairs of modular weights $(n_Q, n_H) = (0, 0), (0, 0.5)$ and $(0.5, 0.5), (0.5, 1)$, respectively, 
and then the mirage unification of soft masses and $A$-terms occurs only for the last pair. 

In these figures, the red colored region represents the 
parameter space where the Higgs boson mass satisfies the current experimental bounds, 
while it is greater than the experimental upper bound in the yellow region.
In the up (s)quark sector that induces a large radiative correction to the Higgs boson mass, 
the particle masses are not so changed even when the value of $\tan \beta$ is not so large 
as taken in Figs.~\ref{higgsM12} and \ref{higgsM34}.
Then we can easily reduce the Higgs boson mass 
by taking a smaller value of $\tan\beta$ in the case the radiative correction is too large. 
Therefore, we recognize the yellow region that can avoid experimental bounds without a difficulty 
by taking a smaller value of $\tan\beta$ than the one adopted in these figures.

The dashed lines stand for the degree of tuning the $\mu$ parameter.
We can see that the tuning becomes milder as the parameter $\alpha_m$ increases,  
and we notice that it can be relaxed above 10$\%$ for $\alpha_m \sim 2$. 

The green and blue lines show the contours with the fixed masses of gluinos and the lightest top squark 
relevant to the Higgs physics we mainly discuss, respectively, 
to the attached values in the unit of GeV. 
The lower bounds of these masses are typically estimated 
as $m_{\tilde{g}} \gtrsim 1.4$ TeV and $m_{\tilde{t}} \gtrsim 700$ GeV.
Thus we find that $m_0$ should be larger than $1.0$ TeV to exceed these bounds.

The colored regions other than those representing Higgs boson mass are 
excluded or disfavored from the other phenomenological reasons.
The EW symmetry breaking cannot occur correctly in the dark gray region 
where $m_{H_u}^2$ doesn't drop down to a small enough value through its RG evolution.
In the brown region, the charged Higgs boson mass is lighter than 400 GeV 
which will induce a too large branching ratio of the $b \rightarrow s \gamma$ process 
mediated by the charged Higgs boson as discussed in Ref.~\cite{Gambino:2001ew}.
The light gray region makes a top squark or a tau slepton LSP. 
The masses of them tend to become smaller as $\alpha_m$ increases and $\mu$ deceases simultaneously.
Therefore, in the region $\alpha_m \gtrsim 1$, LSP is top squark, tau slepton or higgsino depending on the values of the other parameters. 


Let us turn to compare the numerical results obtained from different pairs of the modular weights. 
We can see that the SM-like Higgs boson mass is clearly smaller for $n_Q=0.5$ shown in Fig.~\ref{higgsM34} 
than the one from $n_Q=0$ shown in Fig.~\ref{higgsM12}, 
because a left-right mixing of top squarks are enhanced 
for the latter case with $n_Q=0$ in addition to the large contributions 
from moduli mediation to the top squark masses.  
While it is remarkable that the feature of the degree of tuning the $\mu$ parameter is unchanged 
for the  same value of $m_0$ in the appropriate range of the parameter $\alpha_m$
even if we change the pair of modular weights.
Therefore we conclude that a small $n_Q$ is favored from the observed Higgs boson mass 
and it does not spoil the naturalness of $\mu$ parameter 
that is a strong motivation for the mirage mediation scenario. 

In this scenario, most of the sparticles have almost the same masses roughly equal to $m_0$ 
as indicated in the TeV scale mirage mediation scenario~\cite{Choi:2006xb,Kitano:2005wc}.
Even in this scenario we find different values of modular weights cause the different patterns 
of EW symmetry breaking,
in other words, the Higgs sector which carries $\mu$-parameter considerably depends on 
the choice of modular weights.    
Then the upper bound for the value of $\alpha_m$ to obtain a successful EW symmetry breaking 
depends on the values of modular weights.
When we compare the two cases $(n_Q, n_H) = (0.5, 1)$ and $(0.5, 0.5)$, 
the upper bounds are found as $\alpha_m\sim 2.1$ and  $1.9$ respectively.
Because a larger $n_H$ leads to a smaller up-type Higgs soft scalar mass $m_{H_u}^2$ 
and it is easy to drop down to a negative value through the RG evolution, 
the acceptable value of $\alpha_m$ is raised to 2.1 in the former case.

On the other hand, in the case with the $n_Q=0$, 
relatively heavy squarks bring a large negative contribution to $m_{H_u}^2$ through their RG evolution 
and then upper bound for $\alpha_m$ is larger than 2 even when $n_H$ is small.
Accordingly the value of $\mu$-parameter has to be small 
in order to turn on a successful EW symmetry breaking 
when the value of $\alpha_m$ is close to its upper bound.
Conversely, we should take the allowed maximal value of $\alpha_m$ 
in order to realize a natural value of the  higgsino mass parameter $\mu$ of the order of the $Z$-boson mass.

We exhibit the concrete value of the Higgs boson mass 
and the degree of tuning $\mu$ parameter with appropriate input parameters in Table~\ref{higgsM1234}.
This table tells us that both the allowed Higgs boson mass 
and relaxed tuning of $\mu$ can be realized in the pure mirage mediation model 
with the suitable value of $\alpha_m$ 
which would be determined by some UV physics (e.g. the flux compactification~\cite{Giddings:2001yu}) 
with the enough accuracy.

\begin{figure}[t]
\hfill 
\includegraphics[width=0.45\linewidth]{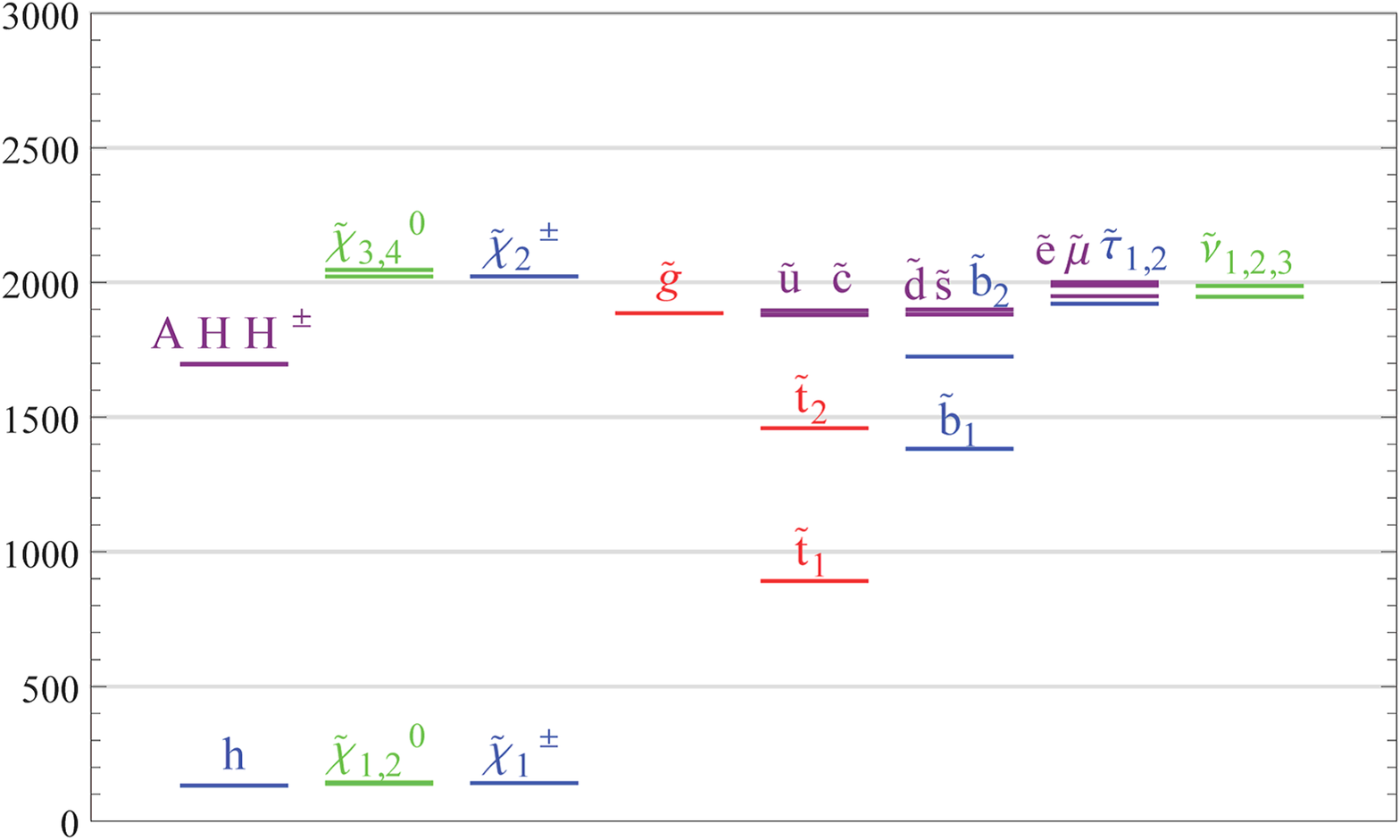}
\hfill 
\includegraphics[width=0.45\linewidth]{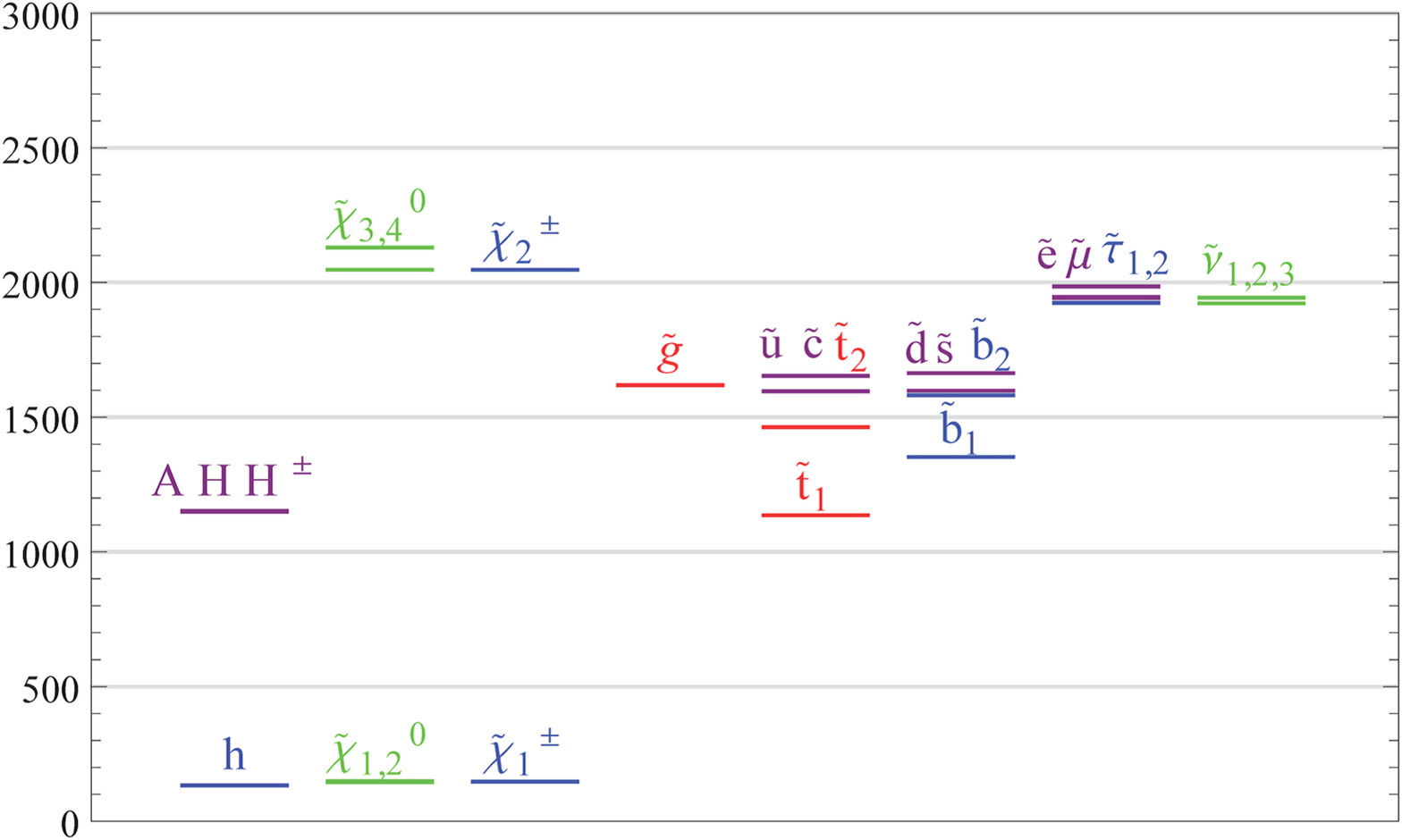}
\hfill 
\caption{The mass spectrum at the sample points M1 (left panel) and M2 (right panel) defined in Table~1. 
The vertical axes measures the mass of each particle in GeV unit. 
The subscript 1 and 2 for the first and the second generation squarks and sleptons are implicit 
in the case that their masses are quite degenerate.}
\label{specM12}
\end{figure}

\begin{figure}[t]
\hfill 
\includegraphics[width=0.45\linewidth]{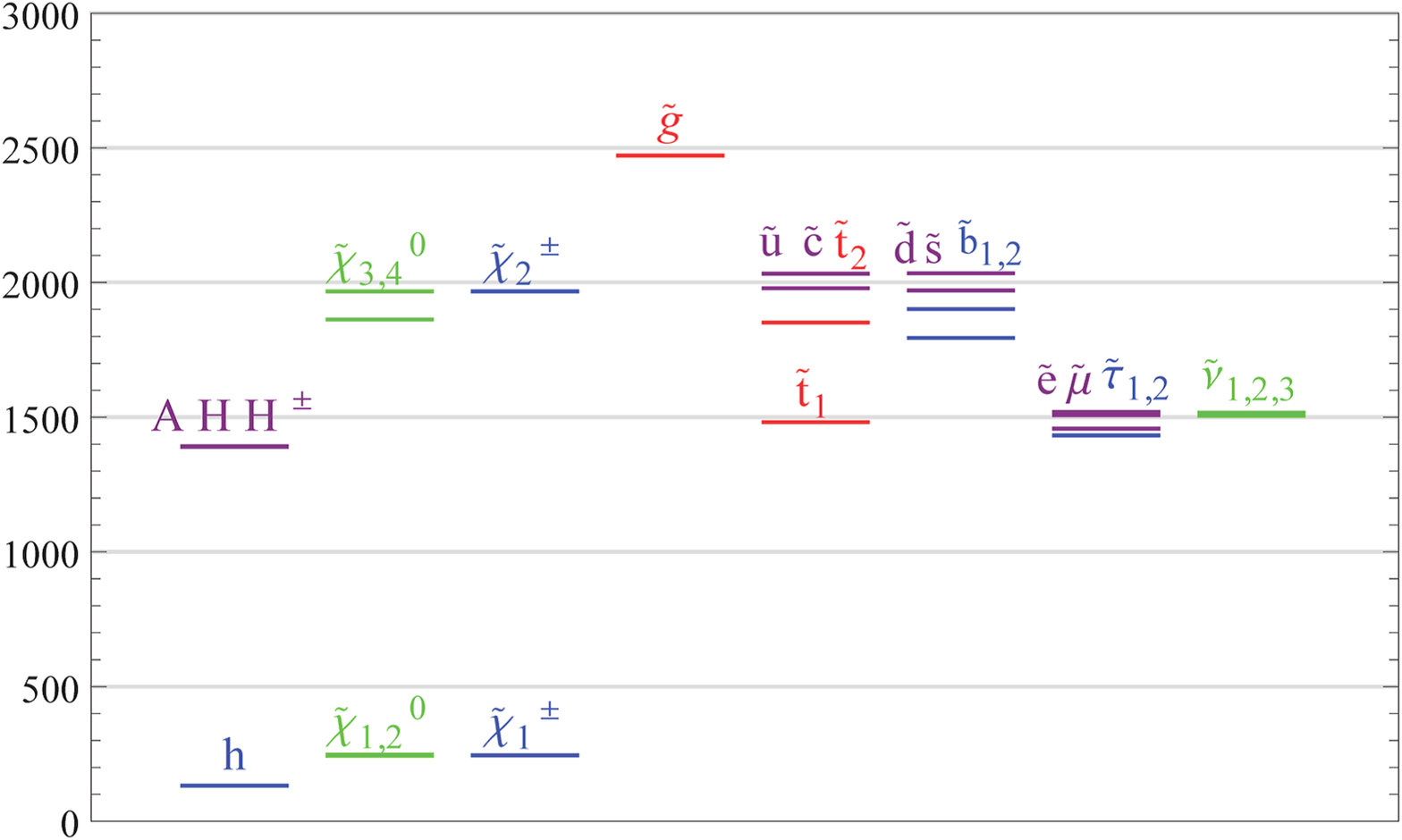}
\hfill 
\includegraphics[width=0.45\linewidth]{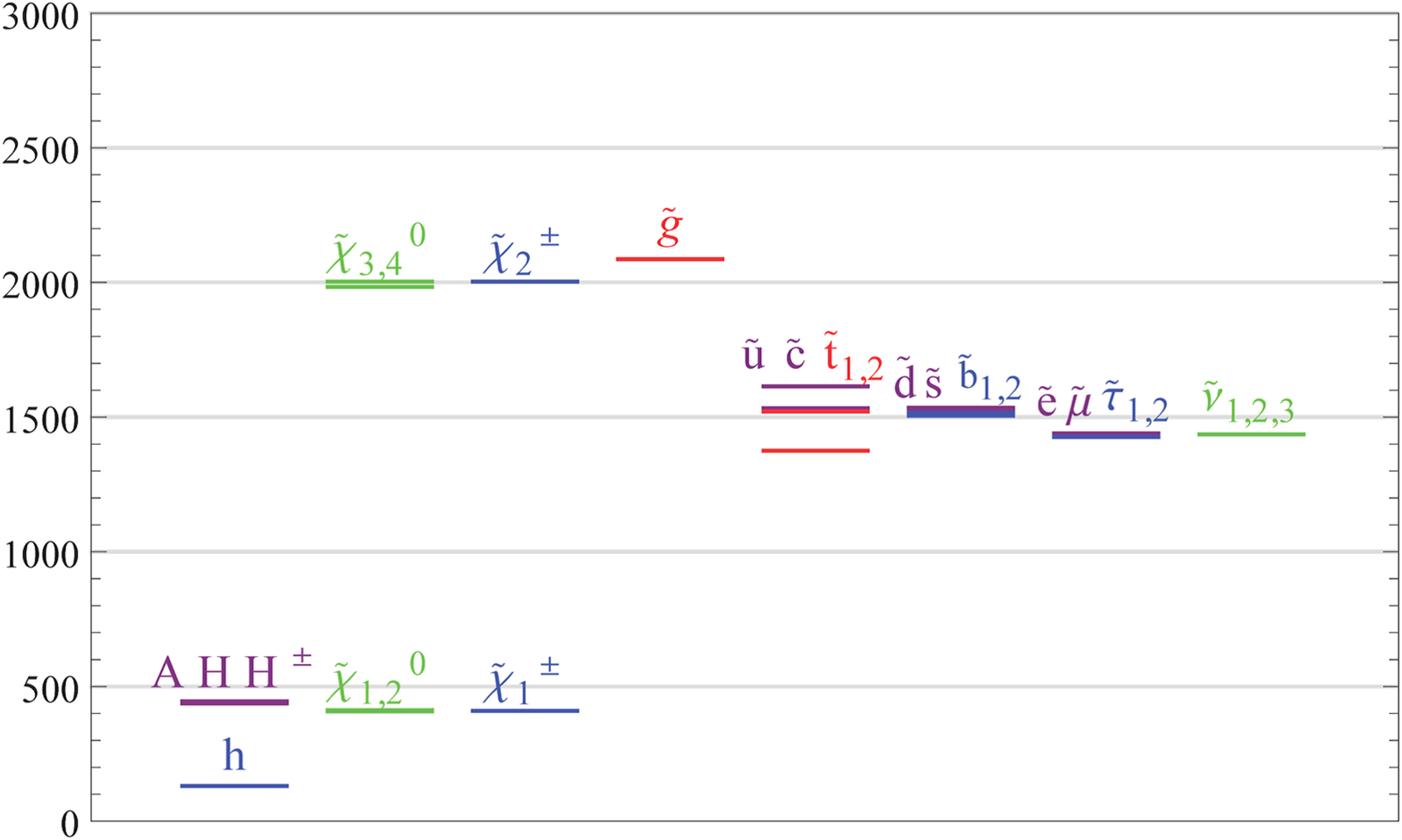}
\hfill 
\caption{The mass spectrum at the sample points M3 (left panel) and M4 (right panel). 
These figures are drawn in the same way as Fig.~3.}
\label{specM34}
\end{figure}

Figs.~\ref{specM12} and \ref{specM34} show the mass spectra at the sample points 
M1-M4 defined in Table~\ref{higgsM1234}.
We can see that experimental lower bounds are satisfied for any sparticle masses 
and a higgsino-like neutralino becomes LSP at the every points.  
The common important feature of these spectra is the almost degenerate gaugino masses 
at the EW scale derived from the TeV-scale mirage condition $\alpha_m\sim 2$ \cite{Choi:2005uz,Endo:2005uy}.
This enhances the large left-right mixing of top squarks 
and at the same time suppresses the RG evolution of the Higgs soft masses due to the cancellation 
between the contributions from gauginos.
As mentioned in Section~\ref{theory}, 
the value of $\alpha_m$ determines the mirage unification scale for the gaugino masses.
The gluino is heavier than the other sparticles for a small $\alpha_m$ yielding a high mirage unification scale, 
which causes heavy squarks because of the RG evolution due to the strong gauge coupling.

In this case, sleptons are lighter than squarks as shown in the left panel of Fig.~\ref{specM34}. 
While, squarks and gluinos are relatively light compared with sleptons and heavy neutralinos, 
respectively, if the mirage unification scale is parametrically lower than the EW scale as shown 
in the right panel of Fig.~\ref{specM34} with a suitable value of  $\alpha_m \lesssim 2$.
Furthermore, heavy (CP-even, odd neutral and charged) Higgs bosons have almost degenerate masses 
and these are light due to the choice of modular weights $(n_Q, n_H) = (0.5, 1)$ 
with which the mirage unification of soft Higgs masses occurs that forces $m_{H_d}^2$ also small.
In this case, the branching ratio of $b \rightarrow s\gamma$ process 
will be enhanced through the diagram mediated by the charged Higgs boson 
and will exceed the experimental bound~\cite{Gambino:2001ew}.
Such a dangerous feature is absent for the other cases of modular weights 
as we can see in Figs.~\ref{specM12} and \ref{specM34}.

\begin{table}[t]
\centering
\begin{tabular}{|c|c|c|c|c|} \hline 
& \multicolumn{4}{|c|}{sample points} \\ \hline\hline
input parameters& M1& M2 & M3 & M4\\ \hline\hline
$(n_Q, n_Q)$ &(0, 0) & (0, 0.5) & (0.5, 0.5) & (0.5, 1) \\ \hline
$\alpha_m$ &2.26&2.42&1.91&2.14 \\ \hline
$m_0$[TeV]& 2.0 &2.0&2.0&2.0 \\ \hline\hline 
output parameters& \multicolumn{4}{|c|}{values}\\ \hline\hline
$100 \times |\Delta_\mu^{-1}|$ (\%) & 55.6 & 28.4 & 7.54 &2.31 \\ \hline
$m_h$[GeV]&125.4&126.2&125.2&123.5 \\ \hline
\end{tabular}
\caption{The mass of SM-like Higgs boson $m_h$ and the degree of tuning $\mu$ parameter, 
$100 \times |\Delta_\mu^{-1}|$ (\%), evaluated at four sample points 
in the parameter space of pure mirage mediation.}
\label{higgsM1234}
\end{table} 
 

\subsection{Deflected mirage mediation}
Next, let us turn to the case with $N_{\rm mess} \neq 0$, 
where the mirage mediation is deflected 
due to the existence of gauge mediated contributions to the soft parameters.
In addition to the parameters of pure mirage mediation, 
there are several ones of its own in the deflected mirage mediation, 
those are the ratio between gauge and anomaly mediated contributions $\alpha_g$, 
the messenger scale $M_{\rm mess}$ and the number of messengers $N_{\rm mess}$.

The messenger scale $M_{\rm mess}$ affects the low-energy mass spectrum through the RG evolution. 
Because the beta functions for the gauge couplings are changed across the messenger scale 
and the size of the gauge couplings increases just above the scale due to the threshold corrections 
induced by the messenger particles.
The number of messengers $N_{\rm mess}$ determines the change of the beta function 
and then it influences not only the size of gauge coupling 
but also the magnitude of gauge mediated contributions to the soft parameters.
In this paper, we analyze the case with $M_{\rm mess}=10^{6}, 10^{12}$ GeV and $N_{\rm mess}=3, 6$.
Among four combinations of $M_{\rm mess}$ and $N_{\rm mess}$, 
the gauge couplings diverge below the GUT scale for one of them, 
$M_{\rm mess}=10^{6}$ GeV and $N_{\rm mess}=6$, 
that contradicts Eq.(\ref{gkf}), i.e., the gauge coupling unification at the GUT scale 
and then we don't treat this combination.
Note that a larger number of messengers $N_{\rm mess}>6$ is inadequate for the same reason.

\begin{figure}
\centering
\hfill 
\includegraphics[width=0.45\linewidth]{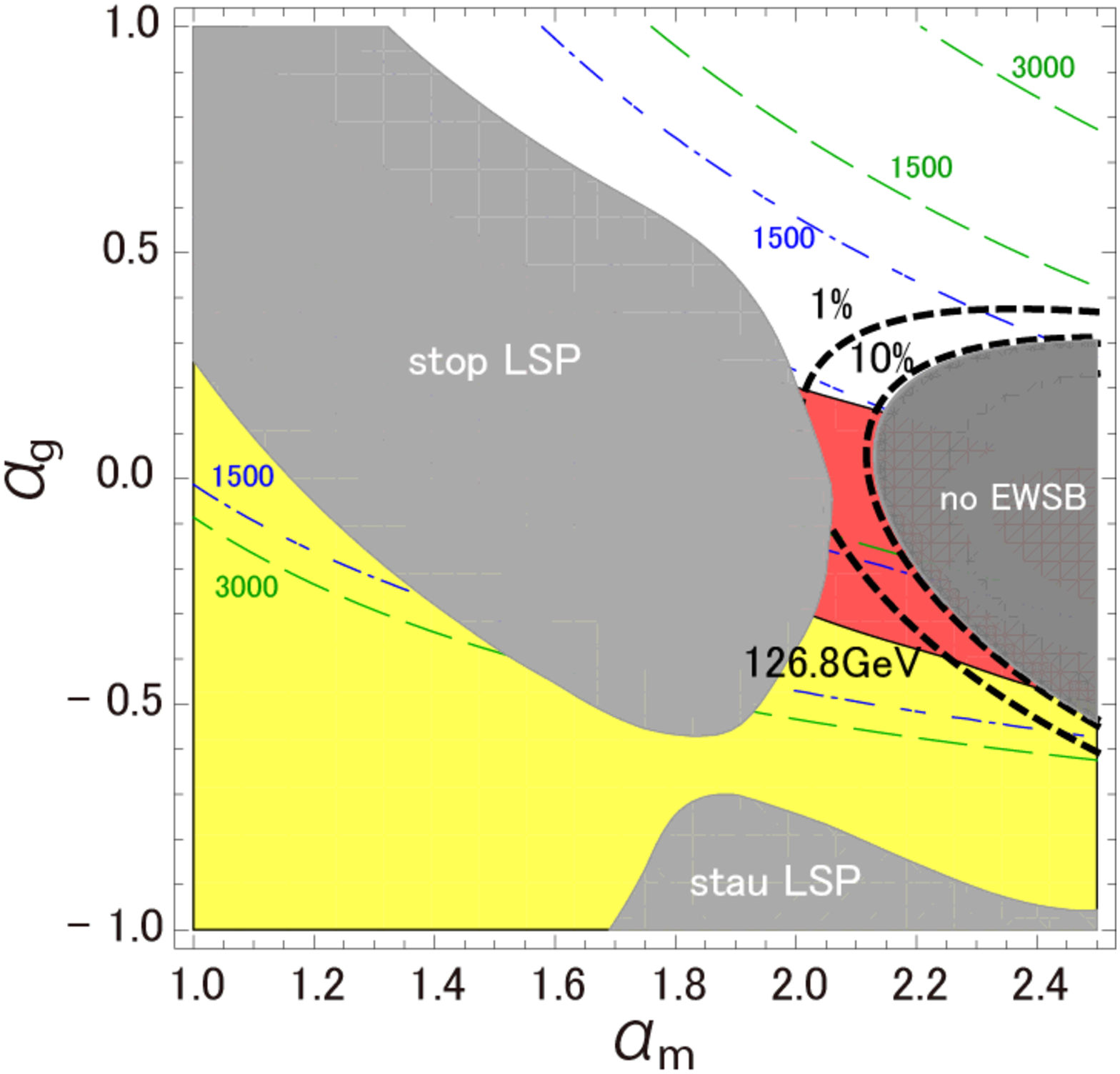}
\hfill 
\includegraphics[width=0.45\linewidth]{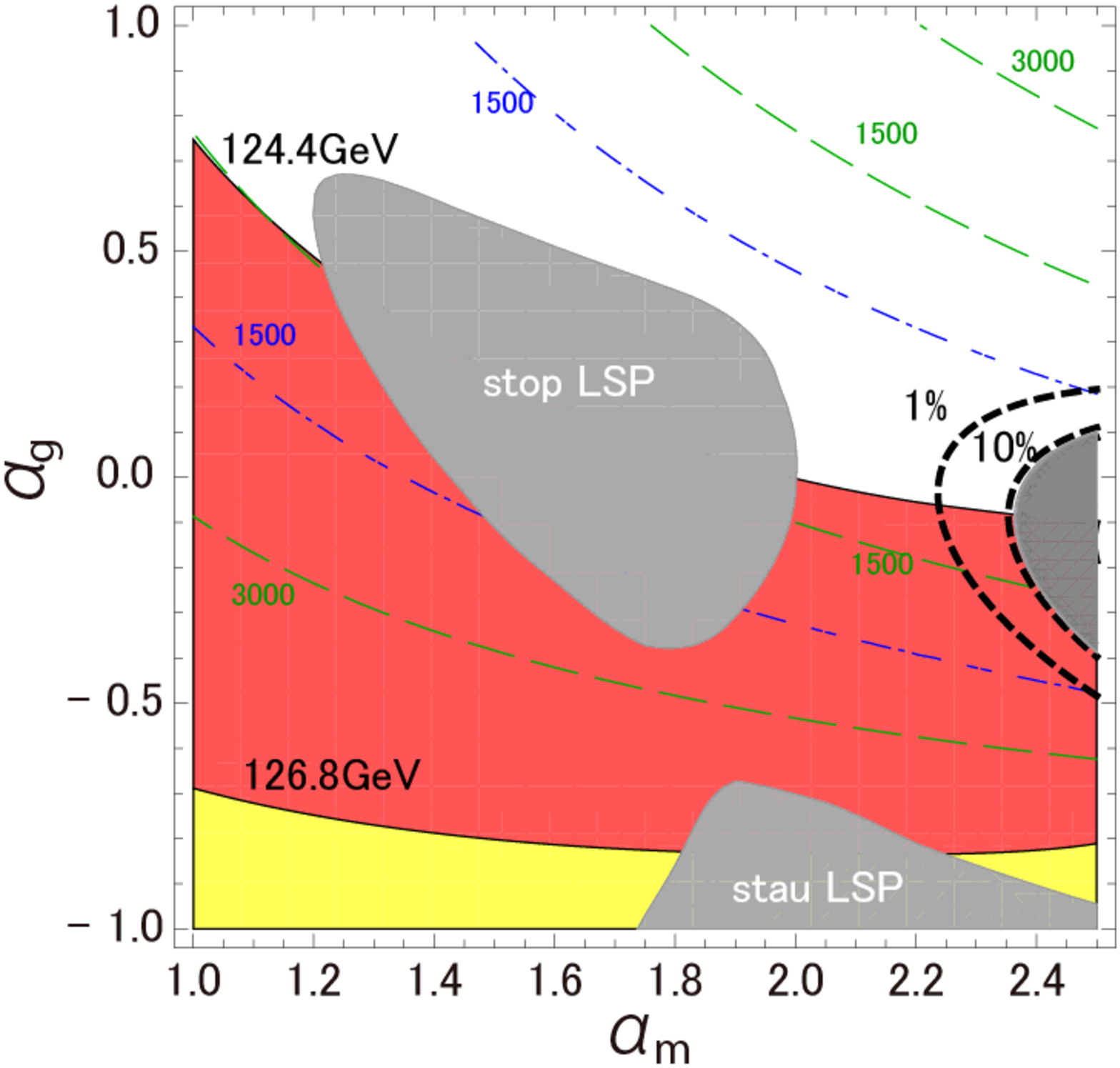}
\hfill 
\caption{Contours of the Higgs boson mass and the degree of tuning  $\mu$ in the deflected mirage mediation 
on $\alpha_m$-$\alpha_g$ plane with the modular weights $(n_Q, n_H) = (0, 0)$ (left panel) 
and $(0, 0.5)$ (right panel)
for $m_0 = 2.0$ TeV, $M_{\rm mess}=10^{12}$ GeV and  $N_{\rm mess}=3$. 
The lines and colored regions are drawn in the same way as those in Fig. \ref{higgsM12}.}
\label{higgsD12}
\end{figure}
\begin{figure}
\centering
\hfill
\includegraphics[width=0.45\linewidth]{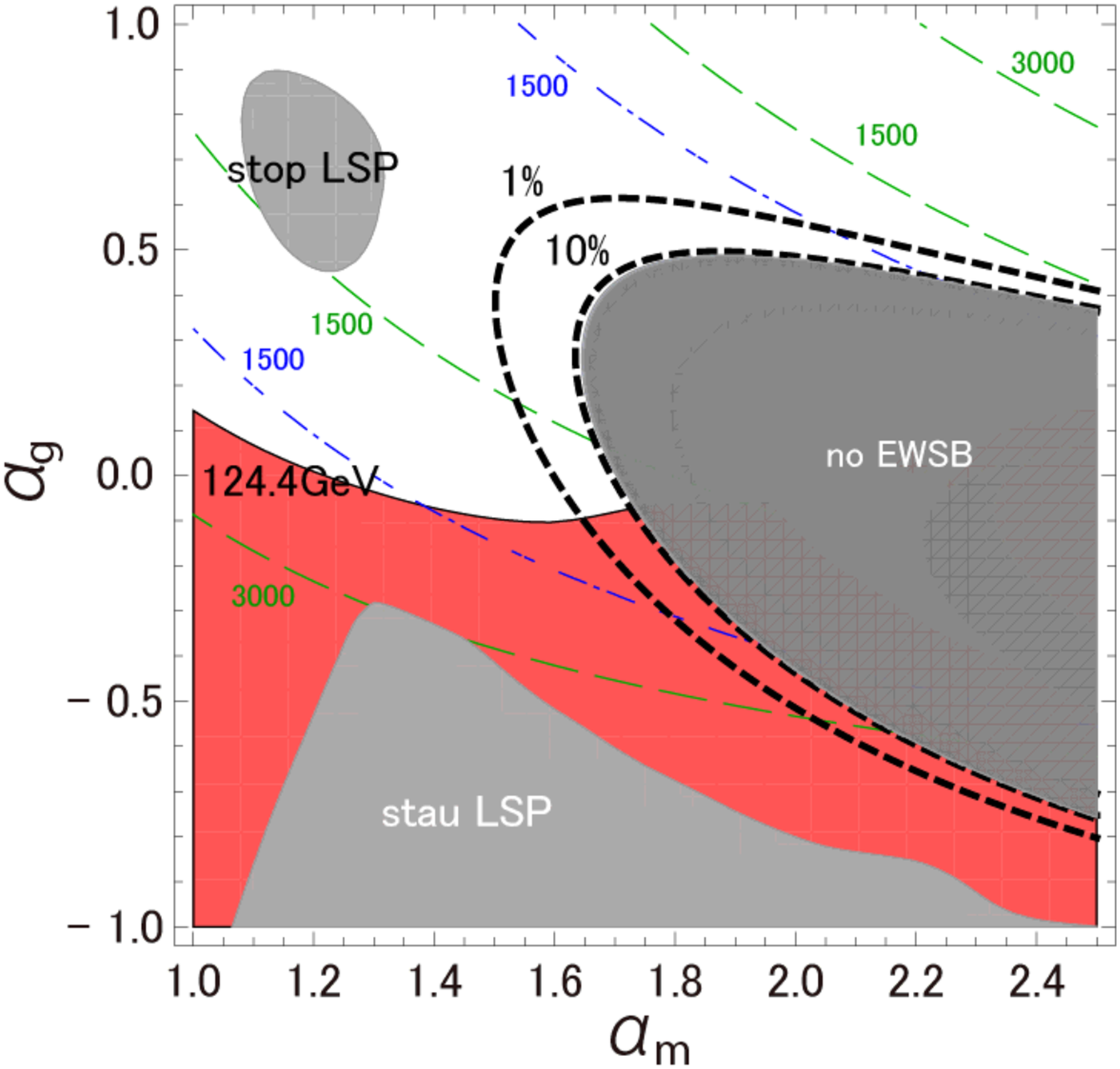}
\hfill 
\includegraphics[width=0.45\linewidth]{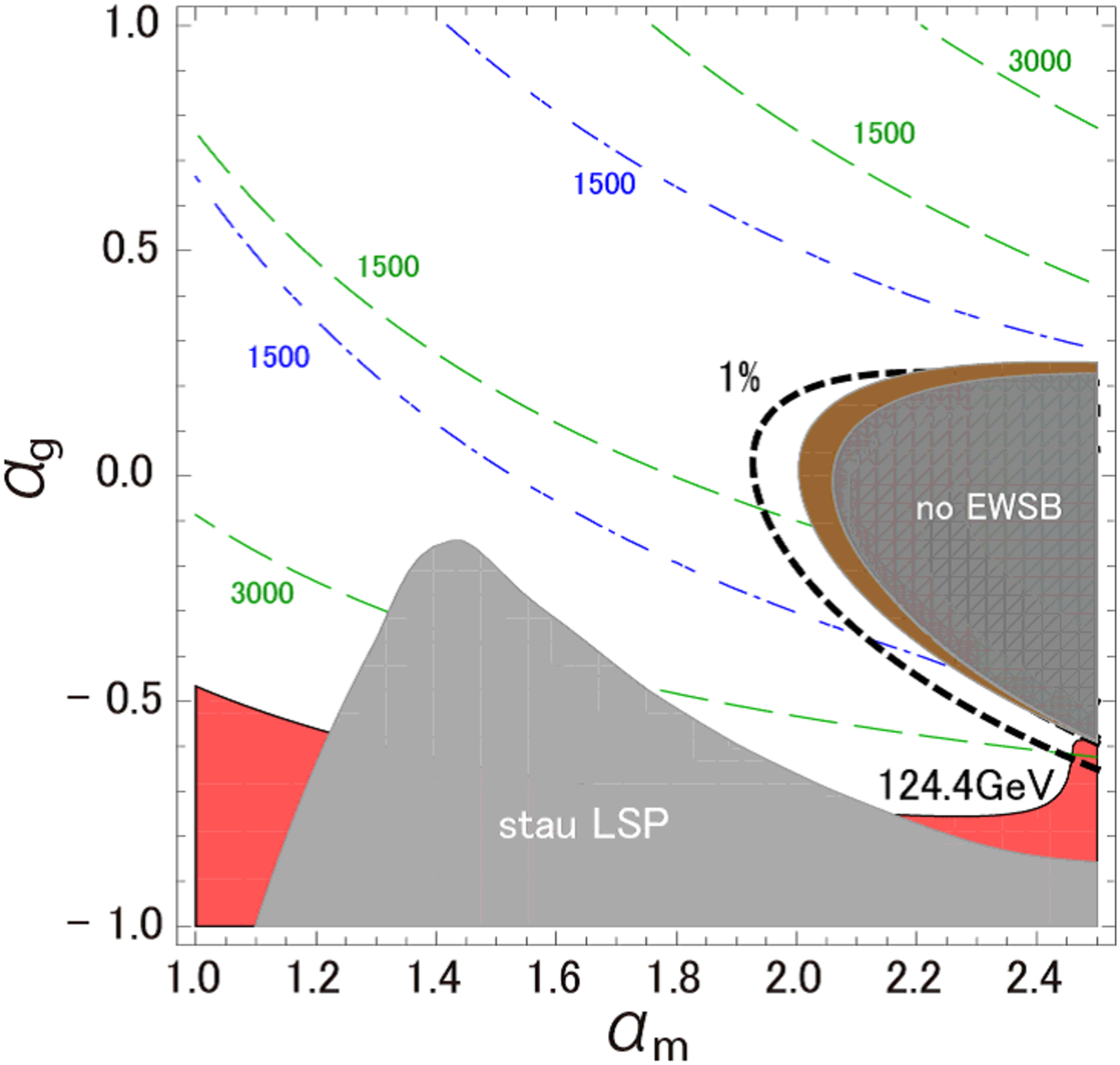}
\hfill
\caption{Contours of the Higgs boson mass and the degree of tuning  $\mu$ in the deflected mirage mediation 
on $\alpha_m$-$\alpha_g$ plane with the modular weights $(n_Q, n_H) = (0.5, 0.5)$ (left panel) 
and $(0.5, 1)$ (right panel) 
for $m_0 = 2.0$ TeV, $M_{\rm mess}=10^{12}$ GeV and  $N_{\rm mess}=3$. 
The lines and colored regions are the same as those in Fig.~\ref{higgsM12}.}
\label{higgsD34}
\end{figure}

\begin{table}[t]
\centering
\begin{tabular}{|c|c|c|c|c|} \hline 
 & \multicolumn{4}{|c|}{sample points} \\ \hline\hline
input parameters& D1& D2 & D3 & D4\\ \hline\hline
$(n_Q, n_H)$ &(0, 0) & (0, 0.5) & (0.5, 0.5) & (0.5, 1) \\ \hline
$(N_{\rm mess}, M_{\rm mess} [{\rm GeV}])$ &(3, $10^{12}$)&(3, $10^{12}$)&(3, $10^{12}$)&(3, $10^{12}$) \\ \hline
$(\alpha_m, \alpha_g)$ &(2.3, -0.35)&(2.4, -0.25)&(1.8, -0.20)&(2.5, -0.60) \\ \hline
$m_0$[TeV]& 2.0 &2.0&2.0&2.0 \\ \hline\hline 
output parameters& \multicolumn{4}{|c|}{values}\\ \hline\hline
$100 \times |\Delta_\mu^{-1}|$ (\%) & 30.9 & 12.1 & 10.6 &4.75 \\ \hline
$m_h$[GeV]&125.7&126.1&124.8&124.5 \\ \hline
\end{tabular}
\caption{The mass of SM-like Higgs boson and the degree of tuning $\mu$ parameter, $100 \times |\Delta_\mu^{-1}|$ (\%), evaluated at four sample points in the parameter space of deflected mirage mediation 
with fixed values of $N_{\rm mess}=3$ and $M_{\rm mess}=10^{12}$ GeV.}
\label{higgsD1234}
\end{table}

\begin{figure}
\hfill 
\includegraphics[width=0.45\linewidth]{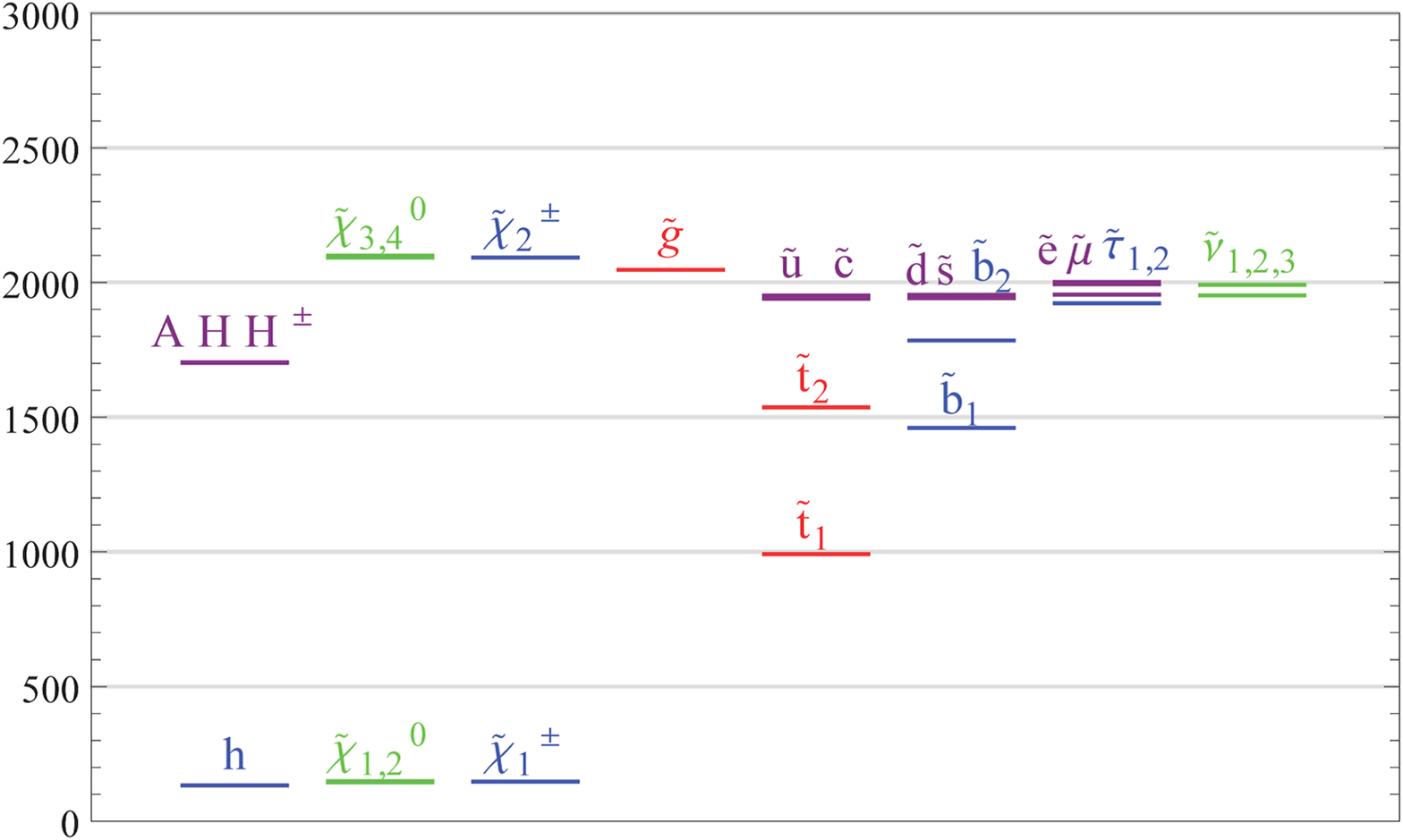}
\hfill 
\includegraphics[width=0.45\linewidth]{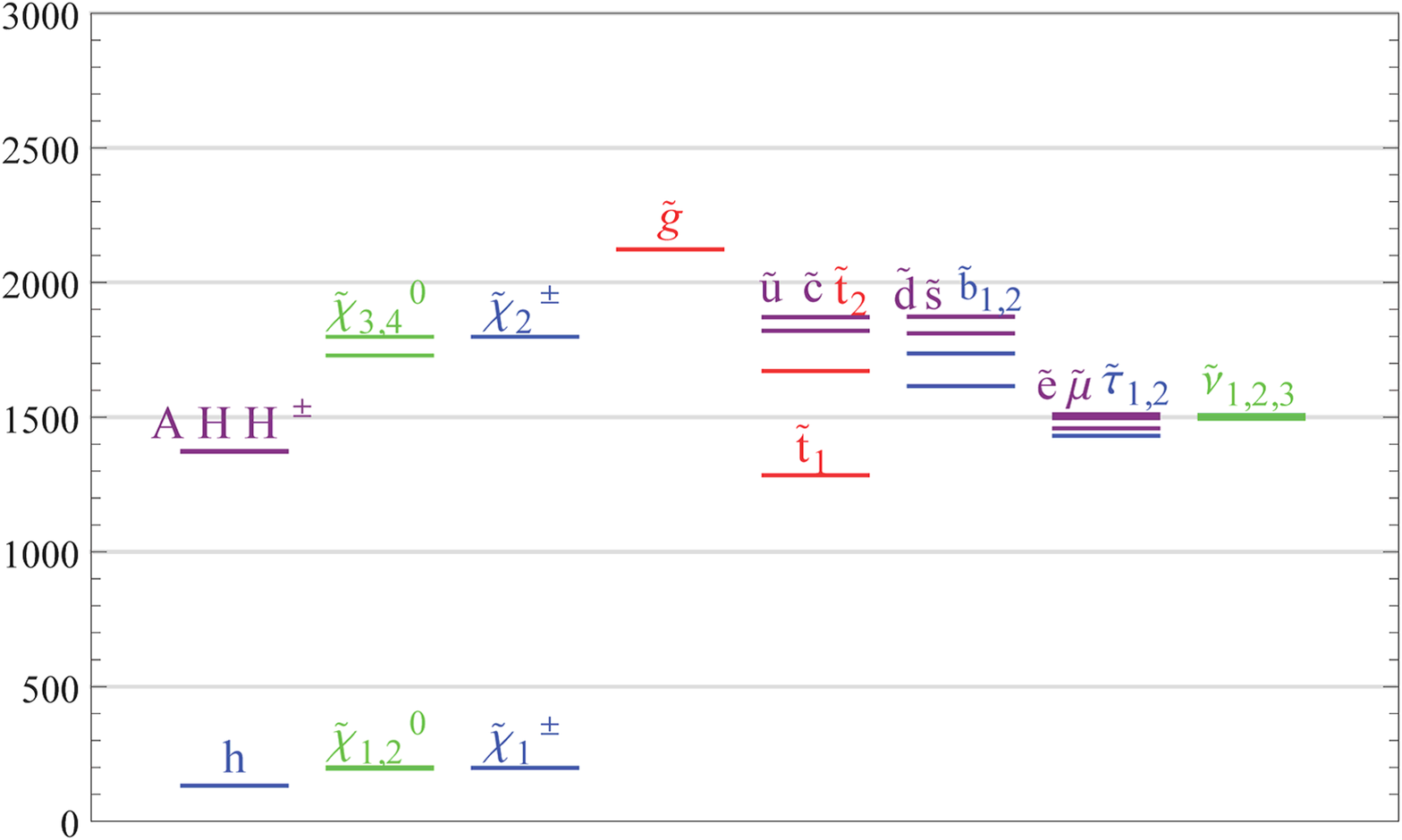}
\hfill 
\caption{The mass spectrum at the sample points D1 (left panel) and D3 (right panel) defined in Table~2. 
These figures are drawn in the same way as Fig.~3.}
\label{specD13}
\end{figure}

Figs.~\ref{higgsD12} and \ref{higgsD34} show the Higgs boson mass 
and the degree of tuning in the deflected mirage mediation, 
which are drawn in the same way as Figs.~\ref{higgsM12} and \ref{higgsM34}, 
for $M_{\rm mess}=10^{12}$ GeV, $N_{\rm mess}=3$.
In Figs.~\ref{higgsD12} and \ref{higgsD34}, the modular weights are chosen 
as $(n_Q,n_H)=(0,0), (0,0.5), (0.5,0.5)$ and $(0.5,1)$ with $m_0=2.0$ TeV.
The Higgs boson mass depends on the modular weights in a similar way to the case of pure mirage mediation,
that is, the left-right mixing of top squarks enhances the Higgs boson mass for the small $n_Q$.
The lightest top squark becomes LSP around $\alpha_g \sim 0$, 
which is caused by the fact that gluino becomes light due to the threshold corrections, 
in this region of $\alpha_g$, 
canceling the original uncorrected mass at the messenger scale. 
On the other hand, $A$-terms for top squarks remain sizable 
around $\alpha_m\sim2$ and $-1 \lesssim \alpha_g \lesssim 0$ 
although the gluino mass is relatively small, 
since the RG effects depending on the initial wino and bino masses push up the size of $A$-terms.
Therefore a large left-right mixing of top squarks is obtained 
around $\alpha_m\sim2$ and $-1 \lesssim \alpha_g \lesssim 0$,  
which keeps the Higgs boson mass large.

Furthermore degree of tuning $\mu$ parameter is relaxed 
in this region, $\alpha_m\sim2$ and $-1 \lesssim \alpha_g \lesssim 0$.
Since the top squark masses increase as $\alpha_g$ departs from the value which minimizes the gluino mass, 
the RG evolution forces $m_{H_u}^2$ to decrease strongly.
Such a gauge-mediated large negative contribution allows the larger value of $\alpha_m$ 
compared with the pure (non-deflected) mirage mediation 
to trigger a collect pattern of EW symmetry breaking.
In other words, we should take a larger value of $\alpha_m$ 
if the contributions form $\alpha_g$ push up the gluino mass, 
in order to realize a natural spectrum with a small $\mu$.
Anyway, we can get a natural spectrum by taking appropriate $\mathcal{O}(1)$ values 
of (would be rational) parameters $\alpha_m$ and $\alpha_g$ 
whenever $-1 \lesssim \alpha_g \lesssim 0$.
Table~\ref{higgsD1234} shows the explicit values of the Higgs boson mass 
and the degree of tuning $\mu$ parameter with the specific input parameters.
We can also construct the models of deflected mirage mediation realizing 
both the observed Higgs boson mass 
and the natural superparticle spectrum by assuming the suitable values of $\alpha_m$ and $\alpha_g$.

Fig.~\ref{specD13} shows the mass spectra with the allowed Higgs boson mass 
and a light higgsino mass at the sample points D1 and D3 defined in Table~\ref{higgsD1234}.
Note that these sample points are selected from the viewpoint of naturalness in such a way 
that the sparticle masses barely exceed the current experimental lower bounds.  
We find that the mass spectra resemble those in the pure mirage mediation.  
Although the basic structure of the mass spectrum mainly depends on the value of $\alpha_m$ 
if we require the small $\mu$ parameter, 
we can realize more abundant patterns of mass spectrum with a nonvanishing $n_Q$ 
due to the existence of gauge mediated contributions.
From the viewpoint of naturalness, $\alpha_m$ should become larger as $\alpha_g$ increases.   
Accordingly, the gluino mass increases rapidly with $\alpha_g$ increasing that leads to heavy squarks.
On the other hand, the slepton masses increase more slowly than the squark ones 
because the masses of wino and bino increase more mildly than that of gluino.
Therefore sleptons can have relatively light masses of  the same order as that of the lightest top squark 
even if the Higgs boson mass is acceptably large while the higgsino remains light for our purpose.

\begin{figure}
 \centering
\hfill 
\includegraphics[width=0.45\linewidth]{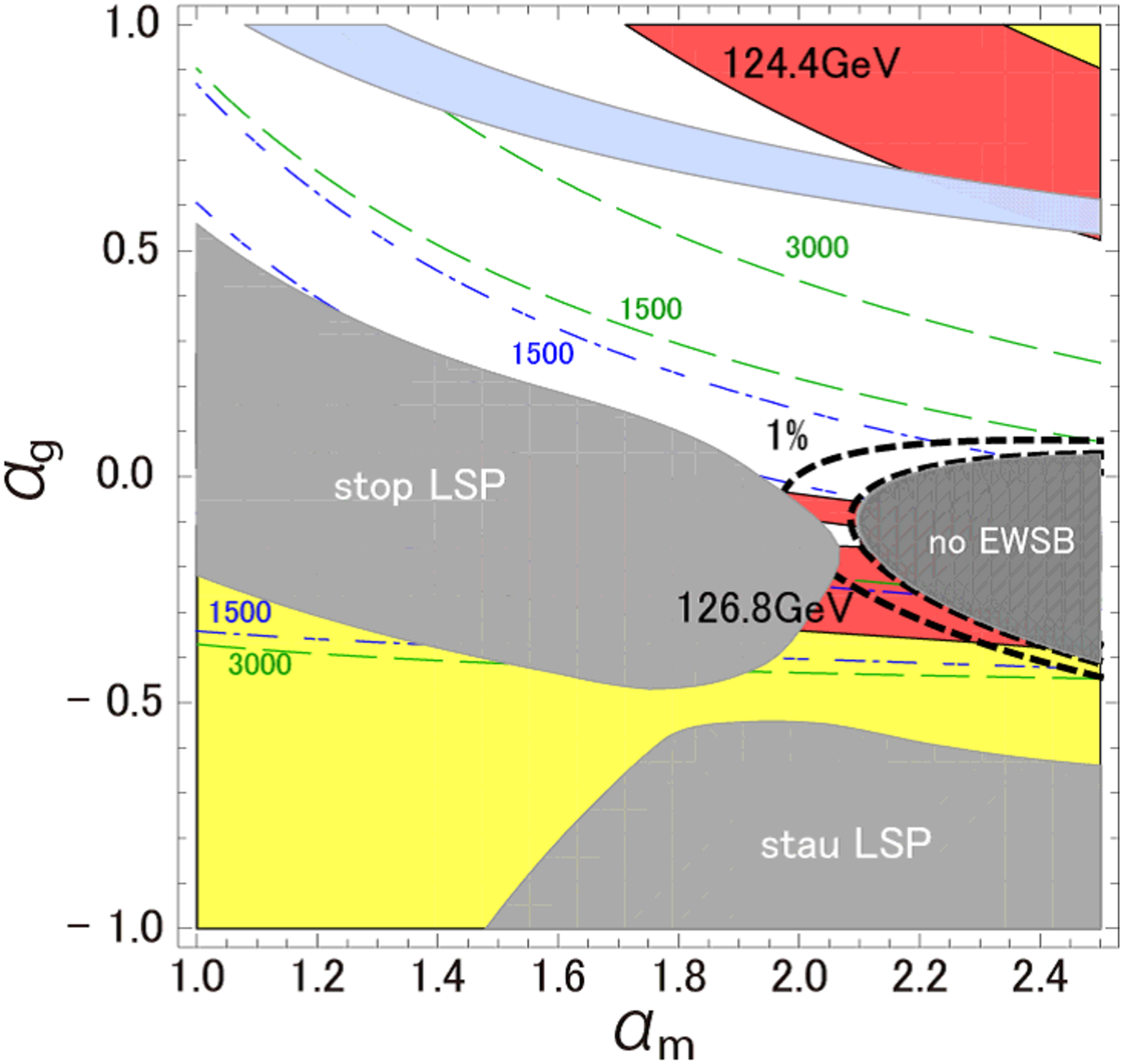}
\hfill 
\includegraphics[width=0.45\linewidth]{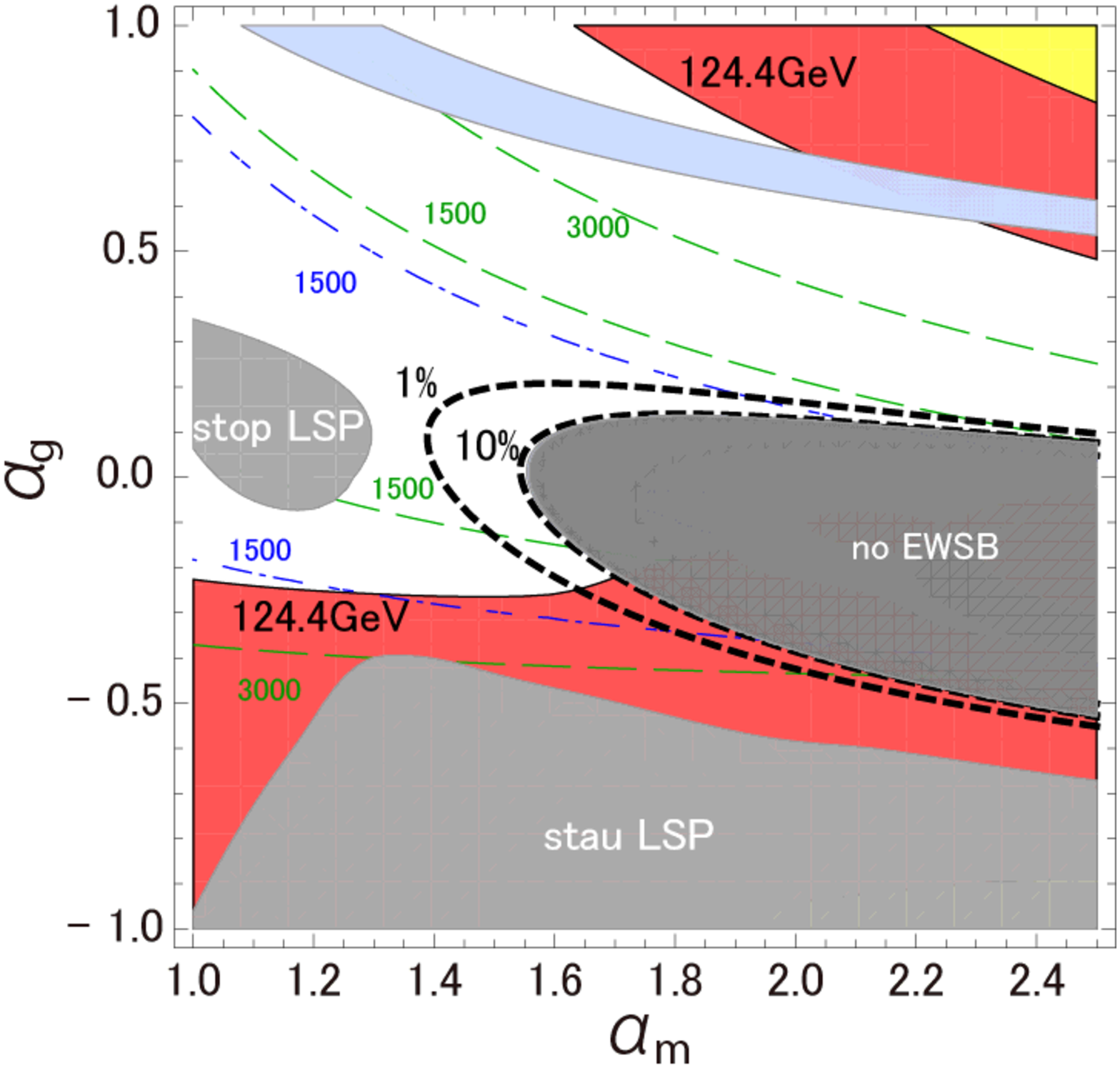}
\hfill 
\caption{Contours of the Higgs boson mass and the degree of tuning in the deflected mirage mediation on 
$\alpha_m$-$\alpha_g$ plane for $M_{\rm mess}=10^{12}$ GeV, $N_{\rm mess}=6$ 
with the modular weights $(n_Q,n_H)=(0, 0)$ (left panel) and  $(0.5, 0.5)$ (right panel) where $m_0 =2.0$ TeV.
The lines and colored regions are drawn in the same way as those in Fig.~\ref{higgsM12}}
\label{higgsD56}
\end{figure}

We also investigated the case with $N_{\rm mess} =6$ and $M_{\rm mess}=10^{12}$ GeV.
The results in this case are shown in Fig.~\ref{higgsD56}.
Due to the larger number of messengers $N_{\rm mess}=6$, 
the value of soft parameters are more sensitive to $\alpha_g$ than the previous case 
and the desired region on $\alpha_m$-$\alpha_g$ plane looks compressed along the $\alpha_g$-direction 
in the figure.
Although the anticipated region in the parameter space is compressed, 
the mass spectrum is not significantly changed compared with the previous cases with smaller $N_{\rm mess}$
including the pure mirage mediation in such a desired region,  
where both the experimentally acceptable Higgs boson mass and the relaxed fine-tuning is accomplished 
with appropriate input parameters as illustrated in Table~\ref{higgsD5678}.  

\begin{figure}
\centering
\hfill 
\includegraphics[width=0.45\linewidth]{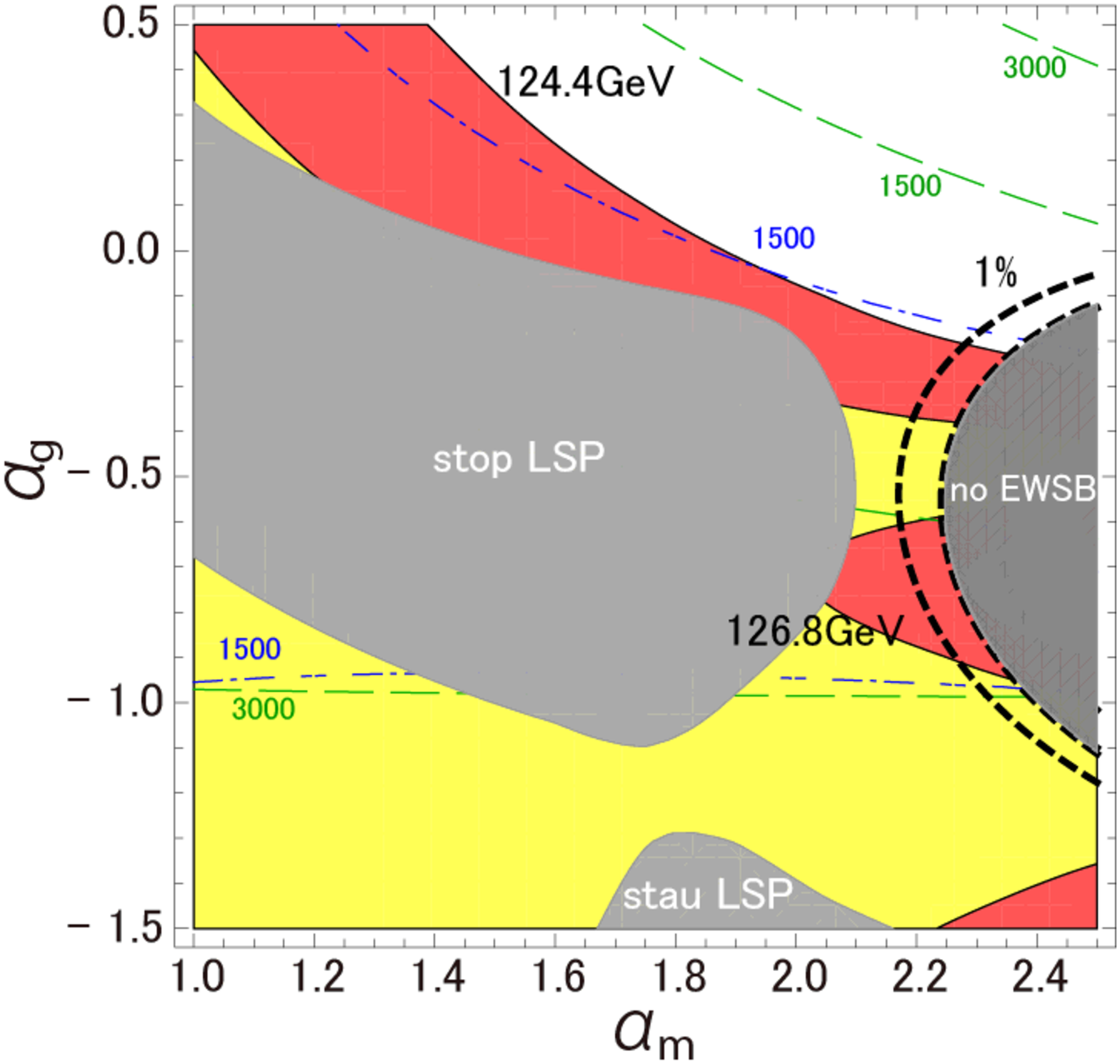}
\hfill 
\includegraphics[width=0.45\linewidth]{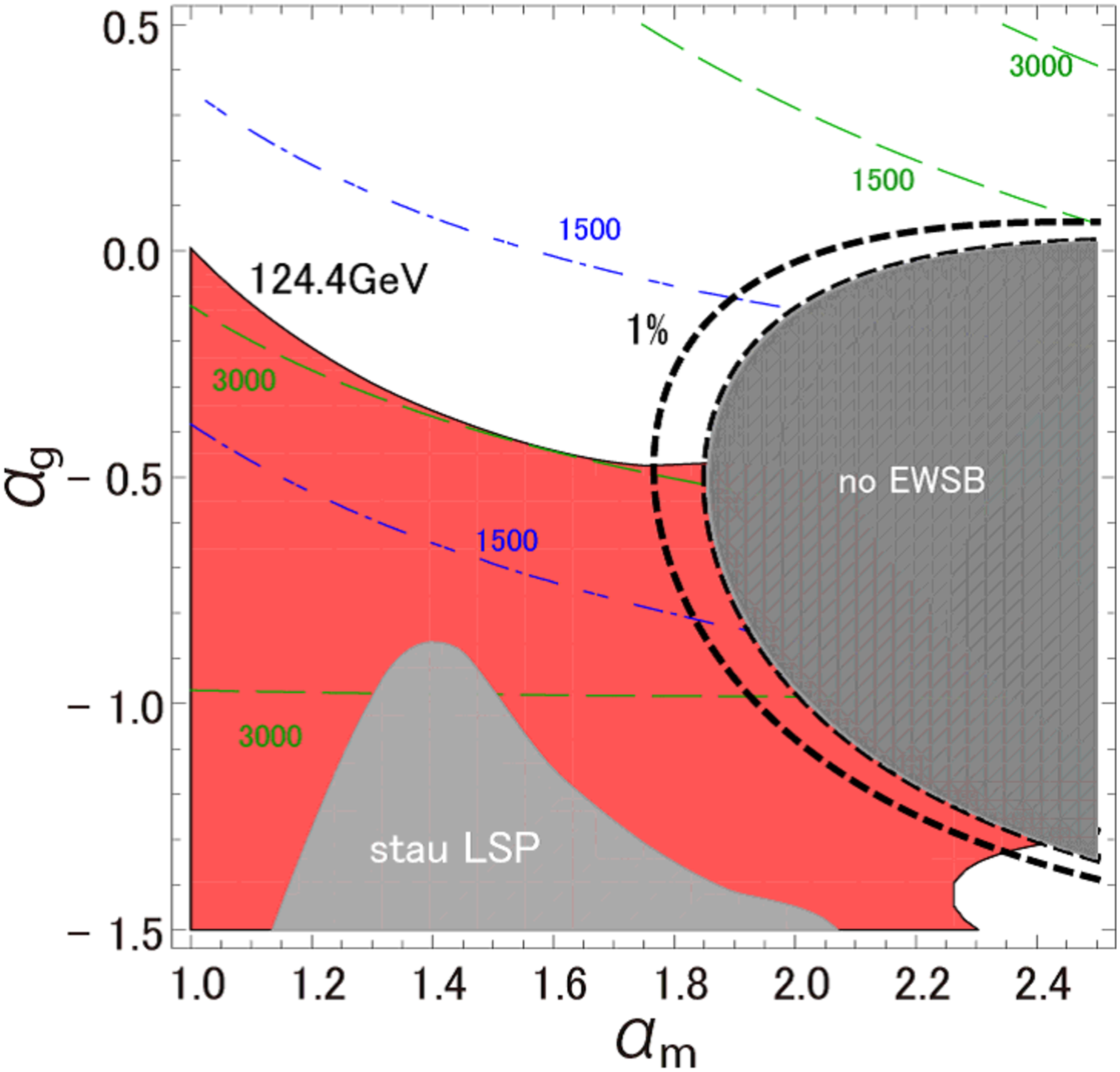}
\hfill 
\caption{Contours of the Higgs boson mass and the degree of tuning in the deflected mirage mediation on 
$\alpha_m$-$\alpha_g$ plane for $M_{\rm mess}=10^{6}$ GeV, $N_{\rm mess}=3$ 
with the modular weights $(n_Q,n_H)=(0, 0)$ (left panel) and  $(0.5, 0.5)$ (right panel) where $m_0 =2.0$ TeV.
The lines and colored regions are drawn in the same way as those in Fig.~\ref{higgsM12}}
\label{higgsD78}
\end{figure}

Finally, in the case of lower messenger scale $M_{\rm mess}=10^6$ GeV with $N_{\rm mess}=3$, 
the aimed parameter region is slightly shifted downward in the $\alpha_g$-direction 
as shown in Fig.~\ref{higgsD78} compared with Figs.~\ref{higgsD12} and \ref{higgsD34}. 
This is because the threshold corrections do  cancel the original uncorrected mass at the messenger scale 
but in the different region of $\alpha_g$ from the one in the previous case with the higher messenger scale 
$M_{\rm mess}=10^{12}$ GeV.
For more detail, the gluino mass is scale invariant above the threshold accidentally 
in the present case with $N_{\rm mess}=3$ and $b'_3=0$.
In addition, the strong gauge coupling has a larger value at the lower messenger scale, 
so the required size of the threshold correction proportional to the strong gauge coupling becomes small.
The RG evolution of the gluino mass is small 
due to the smaller hierarchy between the EW scale and the messenger scale.
Thus negative value $-1 \lesssim \alpha_g \lesssim 0$ brings a light gluino 
which is likely favored from the naturalness.
On the other hand, soft scalar masses are relatively large due to the contributions from the other gaugino masses 
and also the threshold corrections especially in the case with the small modular weights.
Although the favored region is shifted, 
we find again that the sparticle spectra are similar to those in the other cases 
if we restrict ourselves to the desired situation with the experimentally accepted Higgs boson mass 
without tuning the $\mu$ parameter.


\begin{table}
\centering
\begin{tabular}{|c|c|c|c|c|} \hline 
 & \multicolumn{4}{|c|}{sample points} \\ \hline\hline
input parameters& D5 & D6 & D7 & D8\\ \hline\hline
$(n_Q, n_H)$ &$(0, 0)$ & $(0.5, 0.5)$ & $(0, 0)$ & $(0.5, 0.5)$ \\ \hline
$(N_{\rm mess}, M_{\rm mess} [{\rm GeV}])$ &$(6, 10^{12}$)&$(6, 10^{12}$)&$(3, 10^{6}$)&$(3, 10^{6}$) \\ \hline
$(\alpha_m, \alpha_g)$ &$(2.20, -0.26)$&$(1.8, -0.29)$&$(2.25, -0.66)$&$(1.9, -0.77)$ \\ \hline
$m_0$[TeV]& 2.0 &2.0&2.0&2.0 \\ \hline\hline 
output parameters& \multicolumn{4}{|c|}{values}\\ \hline\hline
$100 \times |\Delta_\mu^{-1}|$ (\%) & 9.17 & 6.91 & 11.6 & 18.3 \\ \hline
$m_h$[GeV]&124.5&124.6&125.5&125.4 \\ \hline
\end{tabular}
\caption{The mass of SM-like Higgs boson and the degree of tuning $\mu$ parameter evaluated at four sample points in the parameter space of deflected mirage mediation with various values of $N_{\rm mess}$ and $M_{\rm mess}$.}
\label{higgsD5678}
\end{table}

\subsection{Flavor-dependent mirage mediation}
\label{Fdmirage}
Finally, we study the case with flavor-dependent supersymmetry-breaking mediations 
where squarks and sleptons in the first and the second generations 
are heavier than those belonging to the third generation.
Such a situation could arise when modulus mediation is flavor dependent 
and/or flavor dependent D-term contributions are added to soft parameters.
From a theoretical point of view, the flavor-dependent modulus mediation generally appears, 
e.g., in the case where the Yukawa hierarchy in the SM originates from the wavefunction localization on a cycle
governed by the modulus in extra dimensions 
(see Refs.~\cite{Abe:2004tq} and \cite{Abe:2012fj} for examples 
in five- and ten-dimensional spacetime, respectively, and references therein).
For the latter possibility, D-term contributions to soft scalar masses could arise from the UV-model 
with an anomalous $U(1)_A$ symmetry (see, e.g., Ref.~\cite{Choi:2011xt} and references therein). 
Those depend on flavor indices 
when the anomalous $U(1)_A$ symmetry is identified with a flavor symmetry, 
i.e., the charges for the $U(1)_A$ are flavor dependent 
as those in, e.g., the Froggatt-Nielsen mechanism~\cite{Froggatt:1978nt}.

The most interesting feature of this particular case is that we can avoid the tachyonic sparticles at any scale 
even at the GUT scale.
In the flavor universal (deflected) mirage mediation with the generation-independent modular weights, 
the mass spectrum at a low-energy is consistent with the results of the collider experiments in general.
However squarks and sleptons tend to be tachyonic at around the GUT scale 
especially for $\alpha_m \sim 2$ which is favored from the naturalness.
Therefore some cosmological scenario will be necessary~\cite{Kusenko:1996jn} 
in order to our universe to settle down in the phenomenologically viable vacuum,  
not to drop down into the charge or/and color breaking minima.

Let us consider the soft mass squares at the GUT scale in more detail.
In the (deflected) mirage mediation, these can be decomposed into three parts,  
namely, the contributions from pure modulus mediation, pure anomaly mediation 
and the mixed part of these two.
The pattern of modulus mediation is determined at the tree-level by the modular weights, 
while the other two contributions are arise at the loop-level.
In the purely anomaly mediated part, $\mathcal{O}(y^4)$ terms have positive contributions 
and $\mathcal{O}(y^2 g^2)$ terms give negative ones, 
where $y$ and $g$ represent Yukawa and gauge couplings, respectively.
The terms of $\mathcal{O}(g^4)$ are proportional to the beta functions of corresponding gauge couplings, 
so it has the positive sign for the strong coupling while having the negative sign for the other two gauge couplings.
In the mixed part, 
$\mathcal{O}(g^2)$ terms give the negative contribution to the soft mass squares, 
while $\mathcal{O}(y^2)$ terms have positive sign.

As a result, squarks receive large negative contributions due to the large quadratic Casimir $c_a$ 
appearing in Eq. (\ref{anomalous}). 
Accordingly, their mass squares tend to be tachyonic at around the GUT scale  
in addition to the tachyonic sleptons usually appear in the pure anomaly mediation.
In particular, squarks in the first and the second generations are the most liable to be tachyonic, 
since the positive contributions from $\mathcal{O}(y^4)$ terms contained in the anomaly mediation is quite small 
because of their tiny Yukawa couplings.
Therefore we need a large enough positive contributions from modulus mediation and/or D-term contributions
if we want to avoid the tachyonic scalars even at around the GUT scale 
especially for the first and the second generation squarks.

We first explain that the former case with only modulus-mediated corrections is 
not enough to cure the tachyonic property completely. 
Although some of squarks and sleptons have positive mass squares with $\alpha_m \sim 2$ at the GUT scale 
if the modular weights have small values.
The first and the second generation squarks cannot avoid the tachyonic nature 
in the case of flavor-universal modulus mediation without the large $\mu$ parameter
even if we take $(n_Q,n_H)=(0,0)$ with $\alpha_m \sim 2$.
Therefore we need extra positive contribution to these squark mass squares 
to accomplish both the phenomenologically viable mass spectrum 
without the fine-tuning and the absence of tachyonic particles at any scale below the GUT scale.
However, it is difficult to make the positive squark mass squares at the GUT scale 
with only modulus-mediated contributions 
since they are always accompanied by negative contributions from the mixed parts 
with the anomaly mediation mentioned above.

In order to conclude the arguments to improve the tachyonic properties at the GUT scale, 
we further mention about the other motivations for heavy sparticles in the first and the second generations.
The experimental lower bound for the gluino mass is relaxed when the squarks in these generations are heavy.
A small gluino mass is preferred from the naturalness point of view, 
because the RG evolution of the Higgs soft masses becomes mild, 
while the effects from the first and the second generation squarks are negligible in this evolution.
Another motivation is that hierarchically heavy sparticles in these generations make 
the situation easy to evade unacceptably large flavor-changing neutral currents (FCNCs).
They are hard to mediate the sizable loop diagrams including FCNCs 
especially in the chirality changing processes.

However there are some concerns about the hierarchical sparticle spectra.
Since such a hierarchically heavy soft mass can affect the RG running of the other soft parameters 
even though the corresponding Yukawa couplings are small,
in the case with heavy sparticles in the first and the second generations, 
we should include the contributions from the Yukawa couplings of these generations to keep 
the enough numerical precision. 
Therefore we have calculated the RG flow 
incorporating all the components of Yukawa matrices for completeness throughout this paper. 
The off-diagonal elements in the soft scalar masses are enhanced through the RG evolution 
if the corresponding off-diagonal elements of the Yukawa matrices have sizable values.
These off diagonal elements of soft masses are usually suppressed in the super-CKM basis, 
but these can remain unsuppressed in general if soft masses themselves have hierarchical structures.

In this case we should notice that $m_{H_d}^2$ turns to take a negative value  
through the RG evolution as well as $m_{H_u}^2$ does similarly to the case of larger $\tan \beta$
if the sfermions in the first and the second generations are hierarchically heavy.
It implies conditions for the Higgs potential at the low-energy are hard to be satisfied particularly 
for a small $\mu$-parameter. 
Such a small  $\mu$-parameter tends to be incompatible with the conditions 
for a correct EW symmetry breaking 
such as being stable along the D-flat direction.

From the point of view of FCNCs, 
the flavor-dependent modulus mediation would become a new source of flavor violations 
in the mirage mediation. 
The flavor-dependent parts of the modulus mediation 
don't commute with anomalous dimensions or their derivatives appearing in the anomaly mediated contributions. 
Then it also violates the flavor-blind nature of anomaly mediation at the loop level
in addition to the flavor-violations at the tree-level purely from the flavor-dependent modulus mediation.

\begin{figure}[t]
\centering
\includegraphics[width=0.45\linewidth]{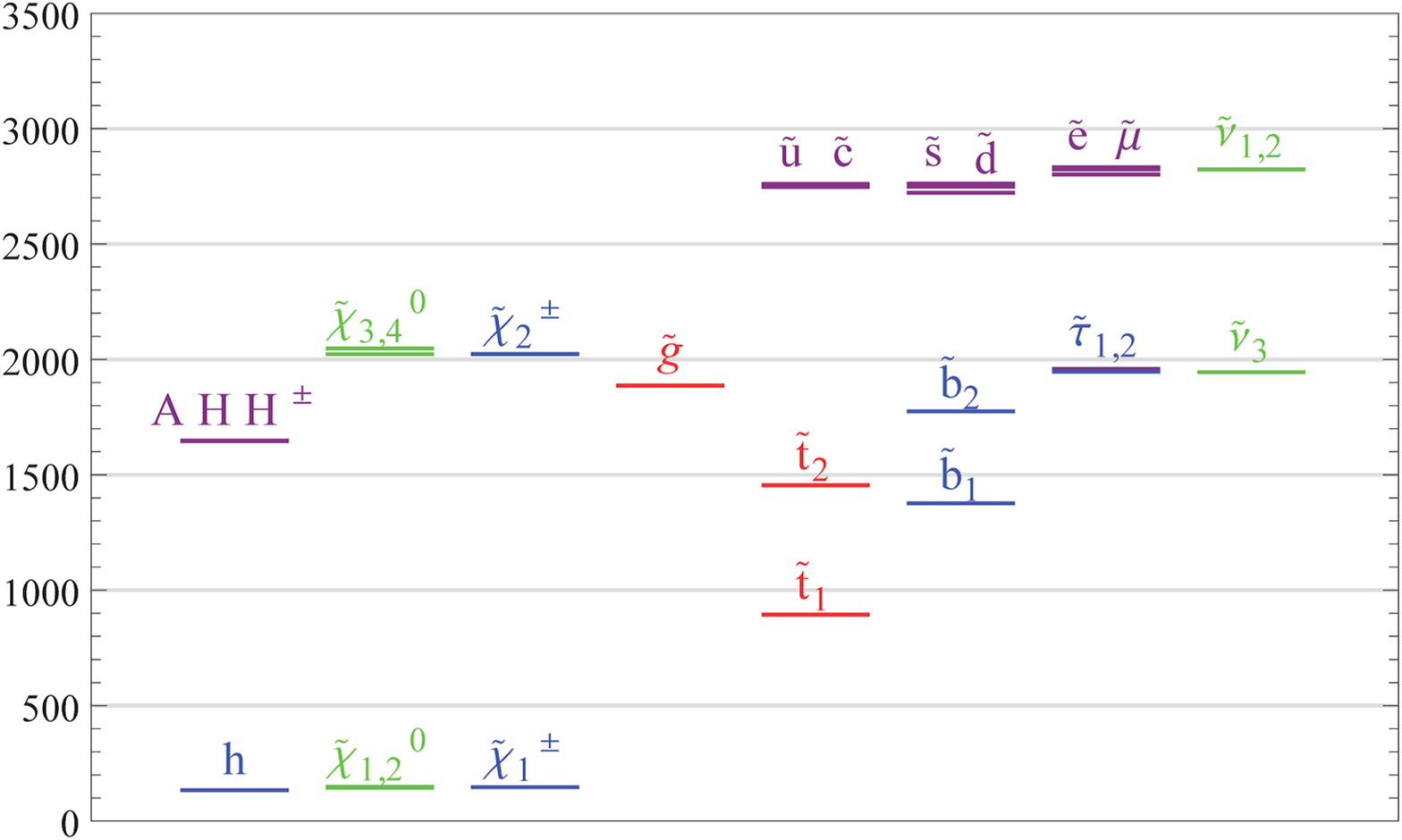}
\caption{The mass spectrum for the case 
that the only soft scalar masses in the first and the second generations receive 
the D-term mediated contributions (\ref{Dterm}) 
in addition to the flavor-universal mirage mediated contributions at the sample point M1 that defined in Table~4}
\label{specT1}
\end{figure}

After all, in this paper, we consider the case with flavor-dependent D-term contributions 
to avoid the tachyonic nature of mirage mediation at the GUT scale. 
We add  the following corrections to the soft masses 
of the first and the second generation ($i,j=1,2$) squarks and sleptons; 
\begin{align}
\label{Dterm}
({\Delta m^2_{\Phi})_i}^j = {\delta_i}^j \times (2.0\ {\rm TeV})^2,
\end{align}
where $\Phi = \tilde{Q}, \tilde{U}, \tilde{D}, \tilde{L}, \tilde{E}$.
In this case, we can avoid the tachyonic sparticles at any scale 
without spoiling the desired structures explained in the previous subsections.
In Table~\ref{specGUT}, we show the values of soft parameters at the GUT scale 
while the masses of Higgs bosons and 
the degree of tuning $\mu$ at the EW scale with specific input parameters.
We can see that all the masses are real-valued and the tachyonic scalars 
whose mass squares are negative are totally absent.
The low-energy spectrum except for the masses of the first and the second generations is virtually 
the same as that of the flavor-universal modulus mediation 
if we compare Fig.~\ref{specT1} with the left panel of Fig.~\ref{specM12}.
It is remarkable that such a non-tachyonic spectrum can be constructed with the help of small modular weights, 
those are also favored from the enhancement of the Higgs mass as mentioned in the previous sections.

Finally, we mention about the tachyonic squark masses at the GUT scale with messengers.
Since the messengers increase the gauge couplings 
while decrease the Yukawa couplings at the GUT scale through the RG evolution,
the soft masses easily become tachyonic even for the third generation sparticles.
As a result, we cannot find the entirely non-tachyonic mass spectrum 
with the messenger fields we have employed in this paper  
if we consider the case $M_{\rm mess} \lesssim 10^{14}$ GeV not to conflict with the approximation 
used in the moduli stabilization 
and to require the sparticle masses exceed the current experimental bounds 
without the fine-tuning at the same time.
\begin{table}
\centering
\begin{tabular}{|c|c|c|c|} \hline 
sparticles&mass [GeV]&(s)particles, parameters & mass [GeV], value \\ \hline\hline
$m_{\tilde{Q}_1}$&980.9&$m_{\tilde{E}_1}$& 582.4 \\ \hline
$m_{\tilde{Q}_2}$&1319&$m_{\tilde{E}_2}$&2059 \\ \hline
$m_{\tilde{Q}_3}$&1334&$m_{\tilde{E}_3}$&2083 \\ \hline
$m_{\tilde{U}_1}$&1769&$M_{H_u}$&3089 \\ \hline
$m_{\tilde{U}_2}$&1775&$M_{H_d}$&738.8 \\ \hline
$m_{\tilde{U}_3}$&2337&$(A_u)_{33}$&3810 \\ \hline
$m_{\tilde{D}_1}$&746.5&$M_{\tilde{B}}$&4968 \\ \hline
$m_{\tilde{D}_2}$&2098&$M_{\tilde{W}}$& 2446 \\ \hline
$m_{\tilde{D}_3}$&2116&$M_{\tilde{g}}$&650.9 \\ \hline
$m_{\tilde{L}_1}$&426.9& $m_h(M_{\rm EW})$&125.3 \\ \hline
$m_{\tilde{L}_2}$&2021&$m_H(M_{\rm EW})$ &1638 \\ \hline
$m_{\tilde{L}_3}$&2021&$100 \times |\Delta_\mu^{-1}| (\%)$&48.32 \\ \hline
\end{tabular}
\caption{The sparticle masses $m_{\tilde{\phi}_i}$, the Higgs masses $M_{H_{u,d}}$, the $A$-term 
for the top squark $(A_u)_{33}$, the gaugino masses $M_{\tilde{G}}$ all at the GUT scale, 
the Higgs masses at the EW scale $m_{h,H}$ and the degree of tuning $\mu$ parameter 
where $M_{H_{u,d}}^2=|\mu|^2+m_{H_{u,d}}^2$. 
The subscripts indicate the mass eigenvalues for left-handed $\tilde{Q}$, 
up-type right-handed $\tilde{U}$, down-type right handed $\tilde{D}$ squarks, 
left-handed $\tilde{L}$, right-handed $\tilde{E}$ sleptons, bino $\tilde{B}$, wino $\tilde{W}$ , gluino $\tilde{g}$, 
CP-even lighter $h$ and heavier $H$ Higgs bosons.
The value of input parameters are the same as the sample point M1: $(n_Q,n_H)=(0, 0)$, $m_0 = 2.0$ TeV and $\alpha_m$=2.26 defined in Table~1.}
\label{specGUT}
\end{table}

\section{Conclusions and discussions}
\label{conclusion}
In this paper, we investigated the mass spectrum in the framework of (deflected) mirage mediation 
particularly focusing on the Higgs boson mass and the higgsino mass parameter $\mu$.
Such a generic framework includes three promising mediation mechanisms 
and then we have treated a fairly general class of hidden supersymmetry breaking scenario.
One of the most important consequences in this paper is 
that two or three types of mediation have to give comparable contributions to the soft parameters,
more precisely $\alpha_m\sim2$ and $-1 < \alpha_g \lesssim 0$, in any case 
if we try to realize the SM-like Higgs boson mass resides in the experimentally allowed region 
without employing an unnaturally large $\mu$ parameter.

Therefore, we need always the comparable anomaly mediations with modulus/gauge mediation 
unless the structure of modulus mediation itself essentially derives our desired property
that strongly depends on details of the models of gravity 
such as supergravity/string models. 
The results in this paper would be quite useful to prove the communication 
with the (low-scale) supersymmetry-breaking sector, 
providing strong suggestions to a model building based on supergravity and superstring theories.

We have found that a small modular weight $n_Q$ for the top quarks is necessary 
to realize the large enough Higgs boson mass owing to the sizable left-right mixing of top squarks 
if we restrict the case such that $m_0 \lesssim 1.5$ TeV.
In this paper, we have assumed only a single modulus remains light 
and affects the low-energy spectrum for simplicity, 
and then controlled the structure of modulus mediation by the modular weights.
It is conceivable that multiple moduli influence the low-energy physics 
and then the modulus mediation have some more structures.
Then a desired sizable left-right mixing may be realized 
due to such a structure of the multiple moduli mediation instead of the small modular weights 
required in the single modulus mediation.

In our framework, the overall size of supersymmetry breaking $m_0$ should be larger than 
about 1.0 TeV to get the experimentally acceptable Higgs boson mass 
even if we take the most suitable values of input parameters.
This conclusion is consistent with the current experimental results from the 
search for supersymmetric particles such as 
$m_{\tilde{t}} \gtrsim 700$ GeV and $m_{\tilde{g}} \gtrsim$ 1.4 TeV.
The typical mass spectrum is that all the sparticles except the higgsino-dominated neutralino 
and the chargino have almost the same masses of the order of $m_0$.
The higgsino mass parameter $\mu$, 
what is crucial for the naturalness argument, can remain small 
by adopting the suitable value of $\alpha_m$ (and $\alpha_g$) 
which would be determined by the moduli stabilization mechanism with a probably enough accuracy.

In more detail, the third generation squarks and sleptons, especially top squarks, 
tend to be lighter than the other squarks and sleptons because of their large Yukawa couplings. 
Thus the candidates for LSP are the top squark, the tau slepton and the higgsino-like neutralinos.
We should take $\alpha_m\sim2$ in order to realize the natural spectrum with the higgsino LSP. 
In this case we may also expect the LSP to play a role of the dark matter~\cite{Choi:2006im, Nagai:2007ud}.
The patterns of gaugino masses depends on the value of $\alpha_m$.  
The gluino mass becomes larger (smaller) than the wino and the bino masses 
as $\alpha_m$ decreases (increases), 
since the mirage unification scale leaves from (approaches to) the EW scale. 
This leads to heavier squarks (sleptons) than sleptons (squarks) as shown in Figs.~\ref{specM12} and \ref{specM34}.

The parameter $\alpha_g$ also influences the gaugino masses depending on the property of gauge mediation 
such as the number of messengers $N_{\rm mess}$ and the messenger scale $M_{\rm mess}$,
thus $\alpha_g$ is important for the RG-evolution of squark and slepton masses too.
However, typical mass spectra are almost identical 
for both the pure mirage mediation and the deflected mirage mediation
with several distinguishable properties of the messenger sector 
if we require both the Higgs boson mass around 126 GeV 
and the degree of tuning $\mu$-parameter relaxed above 10 $\%$.
This typical spectrum can be seen in Figs.~\ref{specM12}, \ref{specM34} and \ref{specD13}.   

We have also proposed and analyzed a model without tachyonic sparticles at any scale below the GUT scale.
Such a situation can be obtained 
only if the value of unified gauge couplings is not so large compared to the case with no messengers 
and also the moduli mediation gives a positive contribution 
sufficient to push the soft scalar mass squares up to positive values.
In this scenario, additional positive contributions to squarks in the first and the second generations are necessary, 
and then we adopted the flavor-dependent D-term contributions (\ref{Dterm}) for such a purpose.
Although these assumptions are required,  there are some more advantages
 in addition to the non-tachyonic mass spectrum at the GUT scale, 
namely, heavy squarks relax the experimental lower bound on the gluino mass 
and also heavy squarks and sleptons could suppress the FCNC processes.

The naturalness argument is the strong guiding principle to construct the model describing new physics.
This requires the MSSM that a particular combination of soft parameters 
should have virtually the same scale as the EW scale 
although the LHC and the other searches have not been discovered any evidence for the new physics.
These suggest a somewhat nontrivial situation should appear 
in the mediation mechanism of supersymmetry breaking as we have discussed in this paper 
that may be explained by a more fundamental theory behind the MSSM.
The LHC would probe such a situation more strictly 
in not so far future that would guide us to construct a more precise description 
of the nature. 

\afterpage{\clearpage}

\subsection*{Acknowledgements}
H.A. was supported in part by the Grant-in-Aid for Scientific Research No. 25800158 from the Ministry of Education, Culture, Sports, Science and Technology (MEXT) in Japan. J.K. was supported in part by a Grant for Excellent Graduate Schools from the MEXT in Japan.

\appendix

\section{Ansatz for the Yukawa matrices} 
\label{yukawa}
The Yukawa matrices at the EW scale we used throughout this paper are chosen for generation indices $i,j=1,2,3$ as
\begin{align}
y^u_{ij} &\simeq 
\left( \begin{array}{ccc}
0.173 \times \epsilon^{5} 
& 0.183\times\epsilon^{3.5}  
& 0.848 \times \epsilon^{2.5} \\
0.258 \times \epsilon^{4}  
& 0.377\times \epsilon^{2.5} 
& 0.379 \times \epsilon^{1.5} \\ 
0.203 \times\epsilon^{2.5} 
& 0.188\times \epsilon^{1} 
& 0.997 \times \epsilon^{0} 
\end{array} \right), 
\nonumber \\
y^d_{ij} &\simeq
\left( \begin{array}{ccc}
0.387 \times\epsilon^{3.5} 
& 0.672\times \epsilon^{4} 
& 0.681 \times\epsilon^{3}  \\ 
0.351 \times \epsilon^{2.5}  
& 0.422\times \epsilon^{3}  
& 0.576 \times \epsilon^{2}  \\ 
0.729 \times\epsilon^{1} 
& 1.07\times\epsilon^{1.5} 
& 0.631 \times \epsilon^{0.5} 
\end{array} \right), 
\nonumber \\
y^e_{ij} &\simeq 
\left( \begin{array}{ccc}
0.186 \times \epsilon^{5} 
& 0.131\times \epsilon^{3} 
& 0.309 \times\epsilon^{3}  \\ 
0.275 \times \epsilon^{4.5} 
&0.702\times\epsilon^{2.5} 
& 0.185 \times \epsilon^{2.5}  \\ 
 0.992\times \epsilon^{3.5} 
&0.998 \times \epsilon^{1.5} 
& 1.04\times \epsilon^{1.5}   
\end{array} \right), 
\nonumber 
\end{align}
where $\epsilon=0.225$ denotes the size of mixing by Cabbibo angle 
and the generation index $i$ ($j$) is contracted with left-handed fields $Q_i$ or $L_i$ 
(right-handed fields $U_j,\ D_j$ or \ $E_j$) depending on the upper scripts $u$, $d$ and $e$. 
These can be constructed from the Froggatt-Nielsen mechanism~\cite{Froggatt:1978nt} 
or quasi-localized matter fields in five-dimensional spacetime~\cite{Abe:2004tq,ArkaniHamed:1999dc, Kaplan:2000av}
and are consistent with observed masses and mixings of quarks and charged leptons at the EW scale.
Of course, these forms are not essential at all for our purpose in this paper 
and we employ them just for numerical concreteness. 


\end{document}